\documentclass[twoside,twocolumn,9pt]{article}
\usepackage{extsizes}
\usepackage[super,sort&compress,comma]{natbib} 
\usepackage[version=3]{mhchem}
\usepackage[left=1.5cm, right=1.5cm, top=1.785cm, bottom=2.0cm]{geometry}
\usepackage{balance}
\usepackage{times,mathptmx}
\usepackage{sectsty}
\usepackage{graphicx} 
\usepackage{lastpage}
\usepackage[format=plain,justification=justified,singlelinecheck=false,font={stretch=1.125,small,sf},labelfont=bf,labelsep=space]{caption}
\usepackage{float}
\usepackage{fancyhdr}
\usepackage{fnpos}
\usepackage[english]{babel}
\addto{\captionsenglish}{%
  \renewcommand{\refname}{Notes and references}
}
\usepackage{array}
\usepackage{droidsans}
\usepackage{charter}
\usepackage[T1]{fontenc}
\usepackage[usenames,dvipsnames]{xcolor}
\usepackage{setspace}
\usepackage[compact]{titlesec}
\usepackage{hyperref}

\usepackage{epstopdf}

\usepackage{mathrsfs, amsmath}
\usepackage{wrapfig}
\usepackage{float}
\usepackage[caption = false]{subfig}

\usepackage{graphicx} 
\graphicspath{{graphics/}} 

\usepackage{lipsum} 

\usepackage{tikz} 

\usepackage[english]{babel} 

\usepackage{enumitem} 
\setlist{nolistsep} 

\usepackage{booktabs} 

\usepackage{subfig}
\usepackage{adjustbox}
\usepackage[utf8]{inputenc}
\usepackage[T1]{fontenc}

\newcommand{\x}{\hat{x}}
\newcommand{\dx}{\mathrm{d}\x}
\newcommand{\brho}{\bar{\varrho}}
\newcommand{\subs}[2]{#1_{\mathrm{\textbf{#2}}} }
\newcommand{\bx}{\mathrm{\textbf{x}}}

\newcommand*\Laplace{\mathop{}\!\mathbin\bigtriangleup}
\newcommand{\bpar}{\bar{\partial}}

\newcommand\blfootnote[1]{%
  \begingroup
  \renewcommand\thefootnote{}\footnote{#1}%
  \addtocounter{footnote}{-1}%
  \endgroup
}

\begin{document}

\twocolumn[
  \begin{@twocolumnfalse}
\vspace{7cm}
\sffamily

\noindent\LARGE{\textbf{Formation of cluster crystals in an ultra-soft potential model on a spherical surface}} \\
\vspace{0.3cm} \\

\noindent\large{Stefano Franzini,$^{\ast}$\textit{$^{a,b}$} Luciano Reatto,\textit{$^{a}$} and Davide Pini\textit{$^{a}$}} \\

\noindent\normalsize{We investigate the formation of cluster crystals with multiply occupied lattice sites on a spherical surface in systems of ultra-soft particles interacting via repulsive, bounded pair potentials. Not all interactions of this kind lead to clustering: we generalize the criterion devised in C.~N.~Likos {\it et al.}, {\it Phys. Rev. E}, 2001, {\bf 63}, 031206 to spherical systems in order to distinguish between cluster forming systems and fluids which display reentrant melting. We use both DFT and Monte Carlo simulations to characterize the behavior of the system, and obtain semi-quantitative agreement between the two. Furthermore, we study the effect of topological frustration on the system due to the sphere curvature by comparing the properties of disclinations, i.e., clusters with fewer than six neighbors, and non-defective clusters. Disclinations are shown to be less stable, contain fewer particles, and be closer to their neighbors than other lattice points: these properties are explained on the basis of geometric and energetic considerations. } \\

 \end{@twocolumnfalse} \vspace{4cm}

  ]

\renewcommand*\rmdefault{bch}\normalfont\upshape
\rmfamily
\section*{}
\vspace{-1cm}


\blfootnote{\textit{$^{a}$~Dipartimento di Fisica ``A.~Pontremoli'', Universit\`a di Milano, 
Via Celoria~16, 20133 Milano, Italy. E-mail: stefano.franzini@sissa.it}}
\blfootnote{\textit{$^{b}$~SISSA, Via Bonomea 265, I-34136 Trieste, Italy. }}




\section{INTRODUCTION}

Clustering is a phenomenon in which particles of a fluid aggregate into mesoscopic structures \cite{Likos_2001}. While it is well known that this behavior can arise from the competition between short-range attractions and long-range repulsions \cite{Sear_1999}, which model depletion and electrostatic forces in colloidal fluids, another class of interactions can also lead to clustering: soft particles interacting through purely repulsive bounded potentials can also form clusters at high density \cite{Mladek_2008}.

In this case, clustering is a cooperative phenomenon where each particle favors complete overlap with few particles over partial overlaps with many particles \cite{Lenz_2012, Pini_2015, Prestipino_2014}. The clusters formed in these fluids form crystals with a number of peculiar properties such as mass transport\cite{Moreno_2007, Coslovich_2011}, and unusual reactions to compression\cite{Mladek_2006, Mladek_2007_bis, Mladek_2008_bis} and shear\cite{Nikoubashman_2012}.

However, not every soft repulsive potential leads to clustering: another predicted phenomenon which can occur in fluids of this class is that of reentrant melting, where crystals only form at low temperatures and melt upon compression. Based on the behavior of the structure factor $S(k)$ in the bulk within the scope of the mean field approximation, Likos et al. proposed a criterion to predict clustering in fluids with bounded, repulsive interactions \cite{Likos_2001}. It states that potentials having power spectra (i.e. Fourier transform) with a negative minimum at $k \neq 0$ lead to the formation of cluster crystals with multiple site occupancy at high density and all temperatures.

More recently Edlund et al. found a similar criterion for patterning in spin systems, showing rigorously that the patterned ground states arise from the presence of a negative minimum in the energy spectrum of the interaction potential \cite{Edlund_2010}.

In this work we investigate the clustering phenomenon in fluids of particles constrained to the surface of a sphere.

In recent years the topic of phase transitions on the sphere has attracted increasing attention because of its relevance as a model for many systems\cite{Guerra_2018, Prauetorious_2018, Lavrentovich_2016, Edlund_2014}. For example, the biological world offers numerous instances of ordered structures on spherical geometries, such as viral capsids \cite{Zandi_2004}, lipid rafts\cite{Jacobson_2007} or pollen patterns\cite{Lavrentovich_2016}. Moreover there is a great interest in self-assembled patterns on spherical surfaces due to the possibility of exploiting such phenomena to manufacture patchy particles avoiding current limitations of the top-down approach \cite{Glotzer_2004, Pawar_2010, Bianchi_2011, Yi_2013}.

Spherical systems not only offer an interesting research topic, but also new challenges. The curvature introduces an additional lengthscale to the system, the radius of the sphere, and the finite size of the surface means that properties of the system depend separately on the surface area and the number of particles it contains, rather than simply on the density\cite{Tarjus_2011}. Moreover, ordered structures are also subject to topological frustration, which is the geometric impossibility of establishing local order everywhere in space: in practice this means that any crystal or vectorial field displays a certain number of irreducible topological defects called disclinations \cite{Tarjus_2011, Garc_a_2013, Nieves_2016}.

We focus our attention on the generalized exponential model of order $4$ (GEM-4), which describes a fluid of ultra-soft particles interacting via a pair potential displaying the clustering behavior mentioned above. Although our investigation is purely theoretical, it comes as a natural question asking which physical system may be described by this model, and could be realized experimentally, at least in principle.

It is not trivial to individuate such a system: in fact, while soft potentials can be used to model the effective interactions between the centers of mass of non compact macromolecules such as polymers, dendrimers or microgels, only some specific molecules display the attitude to form clusters. 

In this respect, an important issue is to which extent ultra-soft pair interactions intended to model the forces between two isolated macromolecules can be trusted at the densities at which clustering is expected to occur. A simulation study \cite{Narros_2010} of an assembly of ring polymers has shown that many-body effects substantially modify the effective pair interaction, to the point that clustering does not take place.     
On the other hand, 
it has been found that flexible amphiphilic dendrimers do allow the formation of cluster crystals, although many-body effects are still very important \cite{Mladek_2008, Lenz_2012, Lenz_2016}. In our case dendrimers of this kind could be grafted to a biological membrane, such as a liposome. It is interesting to note that the realization of such a system is already possible through existing techniques, currently employed in the synthesis of drug carriers with enhanced durability \cite{Tooru_2016, Gao_2013}. However, it is fair to say that, even for an isolated pair of dendrimers, the two-dimensional substrate is expected to affect the form of the effective interaction with respect to that in the bulk, as shown for star polymers confined on a plane \cite{Egorov_2013}. This is true {\em a fortiori} in the present case of a curved substrate, especially when, as in the present study, the size of the effective particles cannot be considered small with respect to the sphere radius. This point has not been investigated here since we are not interested in focusing on a specific physical system, but rather in finding how the spherical topology affects the phase behavior with respect to the extended case for a given interaction model.

The paper is laid out as follows: in Section \ref{ONE} we introduce our model potential for ultra-soft particles on the surface of a sphere. Then, our first objective is to generalize the clustering criterion to spherical systems and describe the density-functional theory used in this study. This is done in Sections \ref{clustering} and \ref{DFT} respectively. In order to test the theoretical predictions, we also performed Monte Carlo simulations according to the procedure briefly explained in Section \ref{simulation}. Our results are presented in Section \ref{THREE}. Specifically, in Section \ref{fluid} we discuss theoretical and simulation results for the homogeneous fluid phase, while in Section \ref{phase} we display the phase diagram of the GEM-4 potential on the sphere, featuring the cluster crystal phases found at high density. A more detailed characterization of the cluster crystals is provided in Section \ref{crystals}, where we focus especially on the differences between disclinations and non-defective clusters by employing both DFT and Monte Carlo simulations of the model. Our conclusions are drawn in Section \ref{conclusions}. Finally, in Appendix~A we give some technical details of the numerical method adopted to minimize the free-energy functional.

\section{MODEL}\label{ONE}

\begin{figure}[t]
\includegraphics[width = 3in]{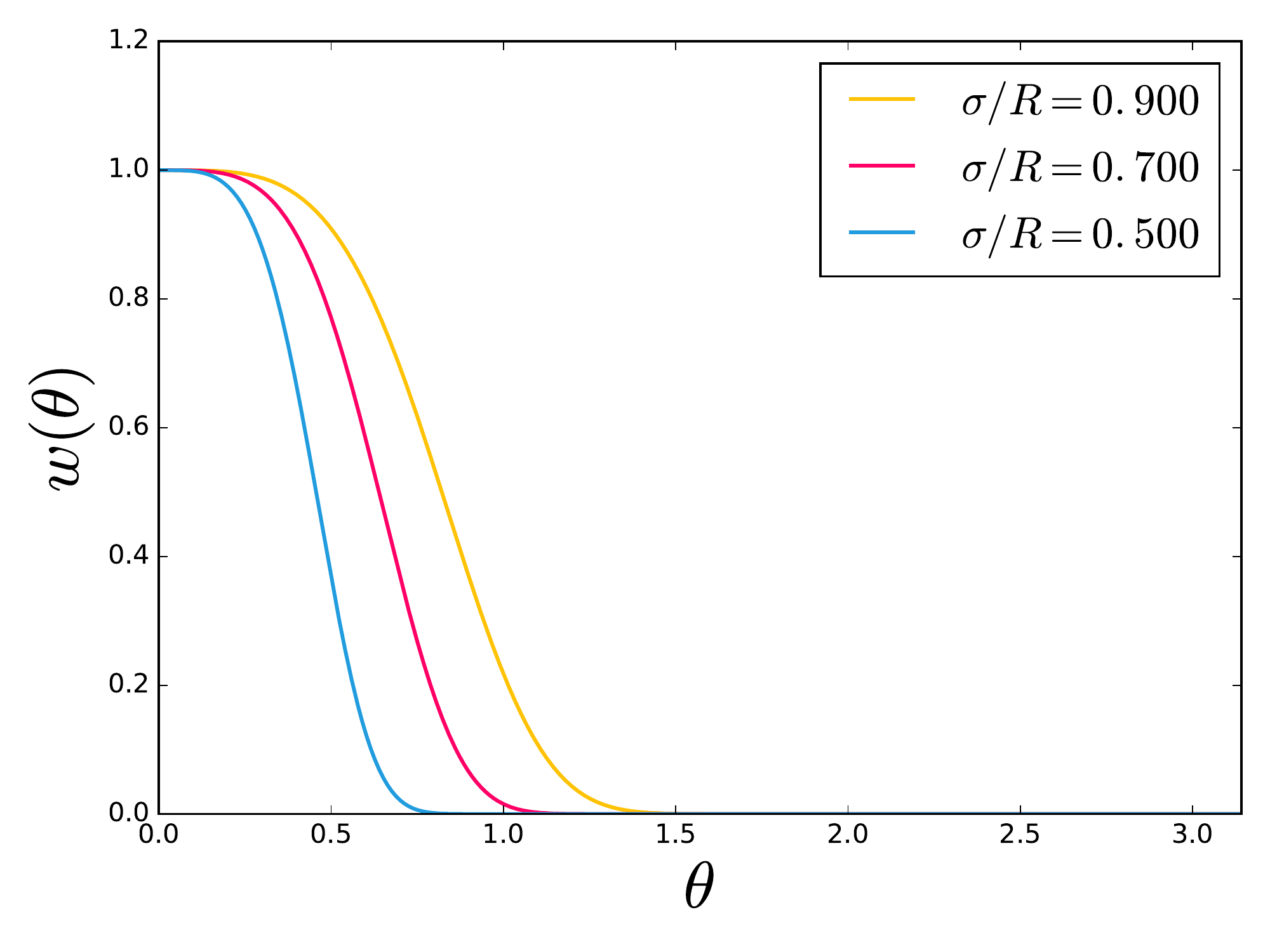}
\caption{The GEM-4 potential for different values of the ratio $\sigma/R$.}
\label{direc_gem4}
\end{figure}

Because of its simplicity and ability to form clusters \cite{Mladek_2007}, we focus on the investigation of the generalized exponential model of order 4 (GEM-4) on the sphere surface. This is a model for colloidal particles constrained to the surface of a sphere and interacting through a bounded, purely repulsive soft pair potential (figure \ref{direc_gem4}) defined by 

\begin{equation}
	w(r) = \epsilon \exp \bigg[ - \bigg(\frac{r}{\sigma}\bigg)^{4} \bigg]
\label{gem}
\end{equation}

Here $\epsilon$ and $\sigma$ define the energy and length scales of the model, and $r$ is the distance between two interacting particles.

This can be used to represent the effective interactions between amphiphilic dendrimers \cite{Mladek_2008}. In this picture, $\sigma$ can be interpreted as the gyration radius of the dendrimers, and the soft repulsive interactions arise from the steric hindrance between monomers of overlapping dendrimers. Because of the entropic origin of the effective interaction, one has $\epsilon\sim k_{\rm B}T$, $T$ being the absolute temperature, and $k_{\rm B}$ the Boltzmann constant, so that the system is actually athermal. However, hereafter we shall follow the lines of references \cite{Mladek_2006, Kahl_2007} in regarding $\epsilon$ as fixed, and taking the dependence on temperature into account. Clearly, this includes the instance of an athermal interaction as a special case.       

Taking into account a system of dendrimers helps us define what distance convention we use: in fact, when considering particles on a sphere, one can either use the usual euclidean distance of the three dimensional space, or the geodesic distance on the sphere (which corresponds to the so called "curved line of force" convention). While it may seem that this has little effect on the model, this choice also has repercussions on the definition of the pressure \cite{Tarjus_2011}. 

In our specific case, the dendrimers spread along the surface of the sphere and cannot penetrate it, so the shortest physical path connecting their centers of mass is given by the great circle passing through them. The length of this curve defines the geodesic distance, which for two points $R\x$ and $R\x'$ on a sphere of radius $R$ can be written as

\begin{equation}
	r = R \cos^{-1}( \x\cdot\x' )
\end{equation}

where $\x$ and $\x'$ are normalized vectors directed along the lines connecting each point with the center 
of the sphere.
Given that $\x\cdot\x' = \cos \theta$, where $\theta$ is the angle between the two points, we may rewrite the potential simply as

\begin{equation}
	w(\x\cdot\x') =
	\epsilon \exp \bigg[ - \bigg( \frac{\arccos(\x\cdot\x')}{\sigma/R}\bigg)^4 \bigg] =
    \epsilon \mathrm{e}^{- \big( \tfrac{\theta}{\sigma/R}\big)^4}
\end{equation}

Here we notice that while in the bulk the only lengthscale is given by the potential range $\sigma$, thus allowing to rescale the system exactly by changing its value, here there is a competition with a second extrinsic lengthscale, the sphere radius $R$, so that the system does not scale exactly with $\sigma$ anymore, and its behavior is controlled by the ratio $\sigma/R$.

To show that this potential leads to clustering on the sphere, in the next section we will obtain a clustering criterion valid on the sphere surface by adapting the argument in \cite{Likos_2001}.

\section{THEORY}\label{TWO}

\subsection{Clustering criterion}\label{clustering}

In this section we establish a criterion for microphase formation on the sphere surface in fluids of particles interacting through soft pair potentials $w(\theta)$. We will set the sphere radius $R=1$ in the following argument.

Following reference \cite{Likos_2001}, potentials of this kind can be divided in two classes depending on the phase behavior of the fluid, namely on whether it displays reentrant melting or clustering at high densities. If the power spectrum displays a negative minimum at a wavenumber $k \neq 0$, then the homogeneous state will be unstable at high densities at all temperatures, and the fluid will form clusters; otherwise, increasing the density will lead to reentrant melting. The former characteristic defines the so called $Q^{\pm}$ class of soft pair potentials.

In the case of a potential defined on the sphere, the power spectrum is given by the coefficients of its expansion over the basis of eigenfunctions which diagonalize the laplacian operator $\Laplace$, which on the sphere are the spherical harmonics $Y_{\ell,m}(\theta,\varphi)$. However, since the potential is isotropic, we can write its expansion using only the zonal spherical harmonics $Y_{\ell}(\theta) \equiv Y_{\ell,m=0}(\theta,\varphi)$, which are also proportional to the Legendre polynomials $P_{\ell}(\cos(\theta))$. We obtain, dropping the index $m$,

\begin{equation}
	w(\theta) = \sum_{\ell} w_{\ell} Y_{\ell}(\theta) \equiv \sum_{\ell} \sqrt{\frac{2\ell+1}{4\pi}}w_{\ell}P_{\ell}(\cos(\theta))
\end{equation}

\begin{figure}[t]
\includegraphics[width = 3in]{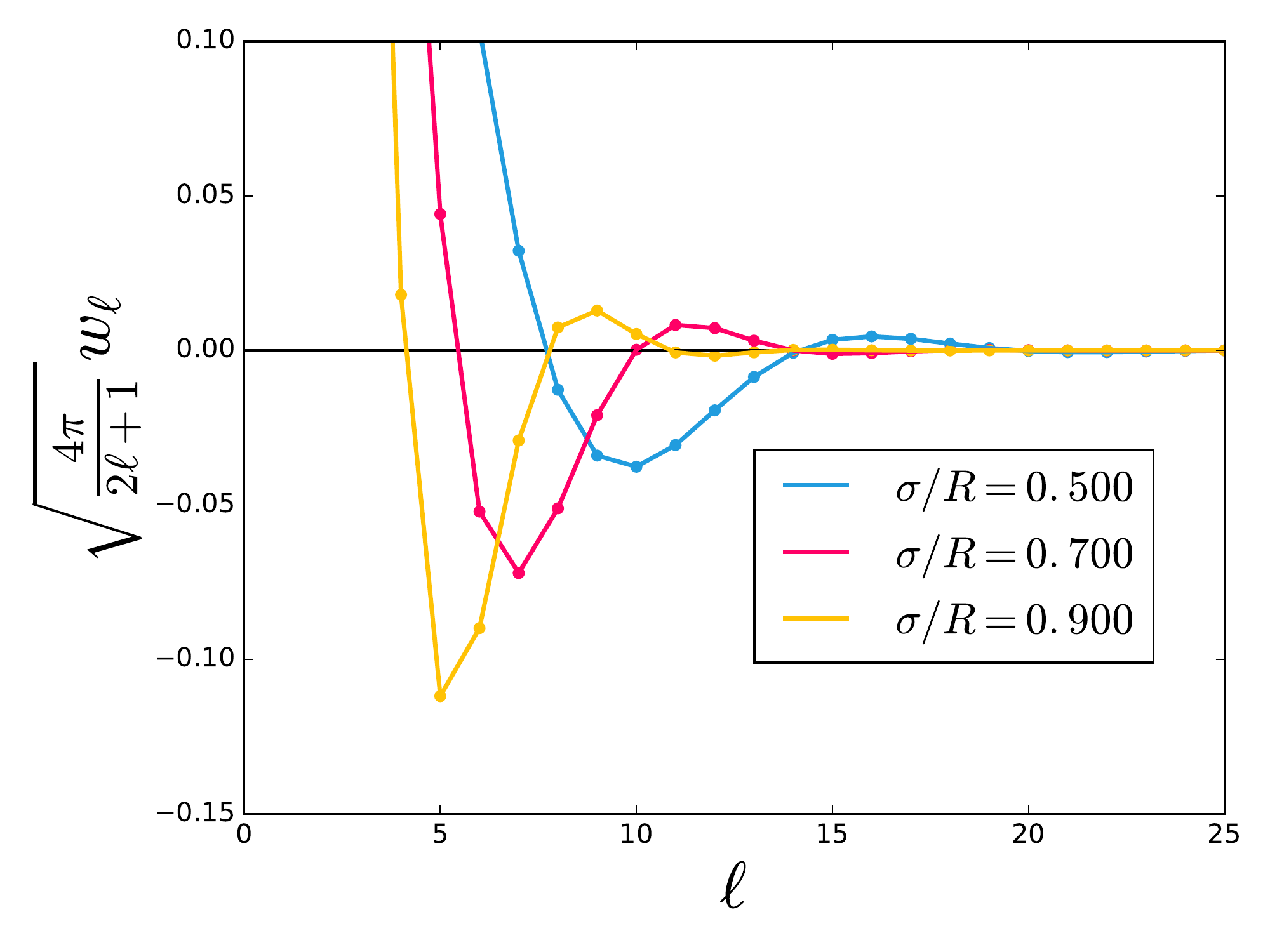}
\caption{Spherical harmonic expansion of the GEM-4 potential for different values of the ratio $\sigma/R$. The harmonic degree $\ell^*$ of the negative minimum increases as $\sigma/R$ decreases.}
\label{spharm_gem4}
\end{figure}

The spectrum of the potential, displaying the peculiar negative minimum which leads to clustering, is shown in figure \ref{spharm_gem4}.

To proceed we treat the homogeneous fluid within the mean field approximation (MFA), which is justified by the theoretical work in reference \cite{Kahl_2007}. The MFA \textit{ansatz} for the direct correlation function $c(\theta)$, which is also isotropic, is given by

\begin{equation}\label{mfa}
	c(\theta) = -\beta w(\theta).
\end{equation}

We use this approximation as a closure of the Ornstein-Zernike (OZ) equation \cite{Temperley_1977} for the total correlation function $h(\x\cdot\x')$, which on the sphere takes the form \cite{Tarjus_2011}

\begin{equation}\label{OrnZern}
	h(\x\cdot\x') = c(\x\cdot\x') + \varrho \int_{S^2} \dx''\, c(\x\cdot\x'')h(\x''\cdot\x')
\end{equation}

this can be solved by expanding the correlation functions on the basis of the laplacian eigenfunctions \cite{Tarjus_2011}, which again are the spherical harmonics. This gives an analytic solution for the coefficients of the total correlation function

\begin{equation}
	h_{\ell} = -\, \frac{\beta w_{\ell}}{1 + \beta\varrho\sqrt{\frac{4\pi}{2\ell+1}} w_{\ell}}
\end{equation}

We can then define an analogous of the structure factor $S_{\ell}$ for a fluid on the surface of a sphere as

\begin{equation}\label{structure}
	\sqrt{\frac{4\pi}{2\ell+1}} S_{\ell} = \frac{1}{1 + \beta\varrho\sqrt{\frac{4\pi}{2\ell+1}} w_{\ell}}
\end{equation}

As in the bulk case, we can encounter two cases depending on the behavior of $w_{\ell}$: (i) $w_{\ell}$ decays monotonically from $w_{\ell=0} = \frac{1}{\sqrt{4\pi}} \int_{S^2} \dx\, w(\x) > 0$ to the value $w_{\ell} = 0$ as $\ell \rightarrow \infty$. (ii) $w_{\ell}$ displays an oscillatory behavior at large $\ell$, which means that it attains negative values in certain intervals. We call $\ell^*$ the harmonic degree at which $w_{\ell}$ has its minimum negative value.

In the latter scenario, $S_{\ell}$ displays a maximum in correspondence to $\ell^*$, which becomes a singularity when

\begin{equation}\label{lambda}
	\varrho = - \sqrt{\frac{2\ell^*+1}{4\pi}} \frac{k_B T}{w_{\ell}}
\end{equation}

By analogy with the bulk case, we call $\lambda$-line the set of points in the phase diagram defined by equation \eqref{lambda}. It delimits the region where the homogeneous phase is unstable.

In fact, the appearance of a singularity in the structure factor signals an instability of the homogeneous phase, since one can interpret $S_{\ell}$ as the response function to an infinitesimal perturbation of harmonic degree $\ell$, on the basis of the fluctuation-dissipation theorem \cite{March_2002}. Hence, the divergence of the structure factor at $\ell^*$ suggests the appearance of an inhomogeneous phase with modulations of characteristic lengthscale $ \sim 2\pi R/\ell^*$.

Notice that in equation \eqref{lambda} an increase in temperature can be compensated by a corresponding increase in density, so that the clustering behavior is present at all temperatures.

Similarly to the clustering criterion obtained here, Edlund et Al.\cite{Edlund_2014} find a correspondence between the presence of a negative minimum in the energy spectrum of the interaction potential in spin systems on the sphere and patterned ground states.

Figure \ref{spharm_gem4} shows the presence of the negative minimum in the spherical harmonic expansion of the GEM-4 potential at $\ell^* \neq 0$. Hence, this potential belongs to the $Q^{\pm}$ class and can lead to the formation of inhomogeneous phases on the sphere.

\subsection{Density-functional theory}\label{DFT}

Having shown that the GEM-4 fluid forms stable inhomogeneous phases above a certain density does not tell us much about the nature of these phases. In order to obtain the phase diagram of the model and to characterize the inhomogeneous phases we employ a simple density-functional theory consisting of a mean field perturbation of a reference fluid, which in our case, since the interactions are bounded, is the ideal gas.

In this picture we write the grand potential functional $\beta\Omega[\varrho]$ as a sum of three terms

\begin{equation}\label{functional}
\begin{aligned}
	\beta \Omega[\varrho] = \beta\mathscr{F}_{id}[\varrho] - \beta\mu N + \beta\mathscr{F}_{ex}[\varrho]\\
	\beta \mathscr{F}_{id} - \beta\mu N = \int_{S^2} \dx\, 
	\varrho(\x)\big[ \ln \big( \varrho(\x)\delta_{\mathit TH}^2  \big) - 1 - \beta\mu \big]\\
	\beta \mathscr{F}_{ex} = \frac{1}{2}\beta \int\int_{S^2} \dx\dx'\, \varrho(\x) \varrho(\x') w(\x\cdot\x'),
\end{aligned}
\end{equation}

where $\delta_{\mathit TH} = \sqrt{\frac{h^2\beta}{2\pi m}}$ is the thermal length on the sphere surface, equivalent to the one found in the bulk bidimensional case.

Notice that for the homogeneous fluid this is equivalent to the mean field approximation used in the previous section. In fact  we can obtain the isotropic direct correlation function $c(\x\cdot\x')$ from its definition \cite{March_2002}

\begin{equation}
	c(\x\cdot\x') = - \lim_{\varrho(\x) \rightarrow \varrho} 
	\frac{\delta^2 \beta\mathscr{F}_{ex}}{\delta \varrho(\x) \delta \varrho(\x')} =
	-\beta w(\x,\x').
\end{equation}

The use of this mean field functional for arbitrary inhomogeneous phases has been justified in reference \cite{Kahl_2007}.

The problem of individuating the stable phase is then reduced to the search of the distribution $\varrho(\x)$ which minimizes the grand potential functional for a given thermodynamic state. These minima must be solutions of the Euler-Lagrange equation obtained by differentiating the grand potential

\begin{equation}\label{euler-lagrange}
	\frac{\delta\beta\Omega}{\delta\varrho(\x)} = \ln\big( \varrho(\x)\delta^2_{\mathit TH} \big) - \beta\mu 
		+ \beta \int_{S^2} \dx' \, \varrho(\x')w(\x\cdot\x') = 0,
\end{equation}

from which we can obtain an exact solution for the homogeneous state by setting $\varrho(\x) = \brho$

\begin{equation}\label{homo_sol}
	\beta\mu = \ln\big(\brho \delta^2_{\mathit TH} \big) + \beta \brho \int_{S^2} \dx' \, w(\x').
\end{equation}

Notice that the homogeneous state is not necessarily a minimum of the grand potential, and in fact it cannot be a minimum beyond the $\lambda$-line, but is always a stationary state, so we can use this solution to parametrize the chemical potential and eliminate the thermal length in equation \eqref{euler-lagrange}

\begin{equation}\label{Gpot}
\begin{aligned}
	\beta \Omega[\varrho] = \int_{S^2} \dx\, \varrho(\x)\big[ \ln \big( \varrho(\x)/\brho \big) - 1 - \beta\mu_{ex} \big] +\\
	+ \frac{1}{2}\beta \int\int_{S^2} \dx\dx'\, \varrho(\x) \varrho(\x') w(\x\cdot\x'),
\end{aligned}
\end{equation}

where $\mu_{ex} =  \brho \int_{S^2} \dx' \, w(\x')$ is the excess chemical potential.

This way the thermodynamic state of the system is specified by setting the values of the temperature $T$, the ratio $\sigma/R$, and the putative density $\brho$. The equilibrium mean density $\varrho$ is then equal to $\brho$ for the homogeneous state, and is expected to be slightly larger in the inhomogeneous phases.

To obtain the density profile of the inhomogeneous phases we need to minimize the grand potential in equation \eqref{Gpot} via numerical methods. We employ the same method used in reference \cite{Pini_2015} adapting it to study spherical systems.

We start by defining a grid of points on the unit sphere, by dividing the altitude $\theta$ over $K$ points and the azimuth $\varphi$ over $2 K$ points. We then sample the density profile at each of these points and use the local densities we obtain as the variational parameters over which we minimize the discretized grand potential $\beta \Omega_D\{\subs{\varrho}{x}\}$ given by 

\begin{equation}\label{Gpot_D}
\begin{aligned}
	\beta \Omega_D \{ \varrho_{\bx} \} = 	
	\frac{\pi^2}{K^2}\sum_{\bx} \sin \theta_{\bx}\varrho_{\bx}
	\big[ \ln \big( \varrho_{\bx}/\brho\big) -1 - \beta\mu^{ex}\big]\\
	+ \frac{\pi^4\beta}{2K^4} \sum_{\bx,\bx'} \sin \theta_{\bx} \sin \theta_{\bx'} 
	\varrho_{\bx} \varrho_{\bx'} w_{\bx,\bx'}
\end{aligned}
\end{equation}

In this study we set $K = 2^8 = 256$, so the total number of sampled is $2^{8} \times 2^9 \sim 130000$. While the total number of sampled points is lower than in reference \cite{Pini_2015}, the lower dimensionality of the systems allows us to obtain double the resolution used there.
To ensure the correctness of the results, the algorithm was also tested using larger values of $K$, with no significant changes in the output density profiles.

The discretized Euler-Lagrange equation becomes a set of equations for the sampled densities

\begin{equation}\label{EL_D}
	\ln\big( \subs{\varrho}{x}/\brho\big) - \beta\mu^{ex} + 
	\frac{\pi^2}{K^2}\beta \sum_{\bx'} \sin \theta_{\bx'}\, \varrho_{\bx'} w_{\bx,\bx'} = 0.
\end{equation}

To obtain a local solution of these equations we start from a trial density profile and apply the preconditioned conjugate gradients algorithm with adaptive step size devised in reference \cite{Pini_2015}. 

Clearly, there is no guarantee that such a local solution is also the global minimum of the grand potential. However, we do perform a rather thorough search for the global minimum by applying this method multiple ( $\sim 10$) times to different starting density profiles for each state, and comparing the grand potential values of the solutions. Moreover, random perturbations are added during the minimization to help explore the grand potential landscape. The details of the algorithm are given in Appendix~A.

Here we limit ourselves to pointing out the main difference from the algorithm applied to bulk systems in reference \cite{Pini_2015}, namely the usage of spherical harmonic expansions instead of Fourier series to get rid of the cumbersome double summations in equations \eqref{Gpot_D} and \eqref{EL_D}.

\begin{figure}[t!]
\includegraphics[width = 3in]{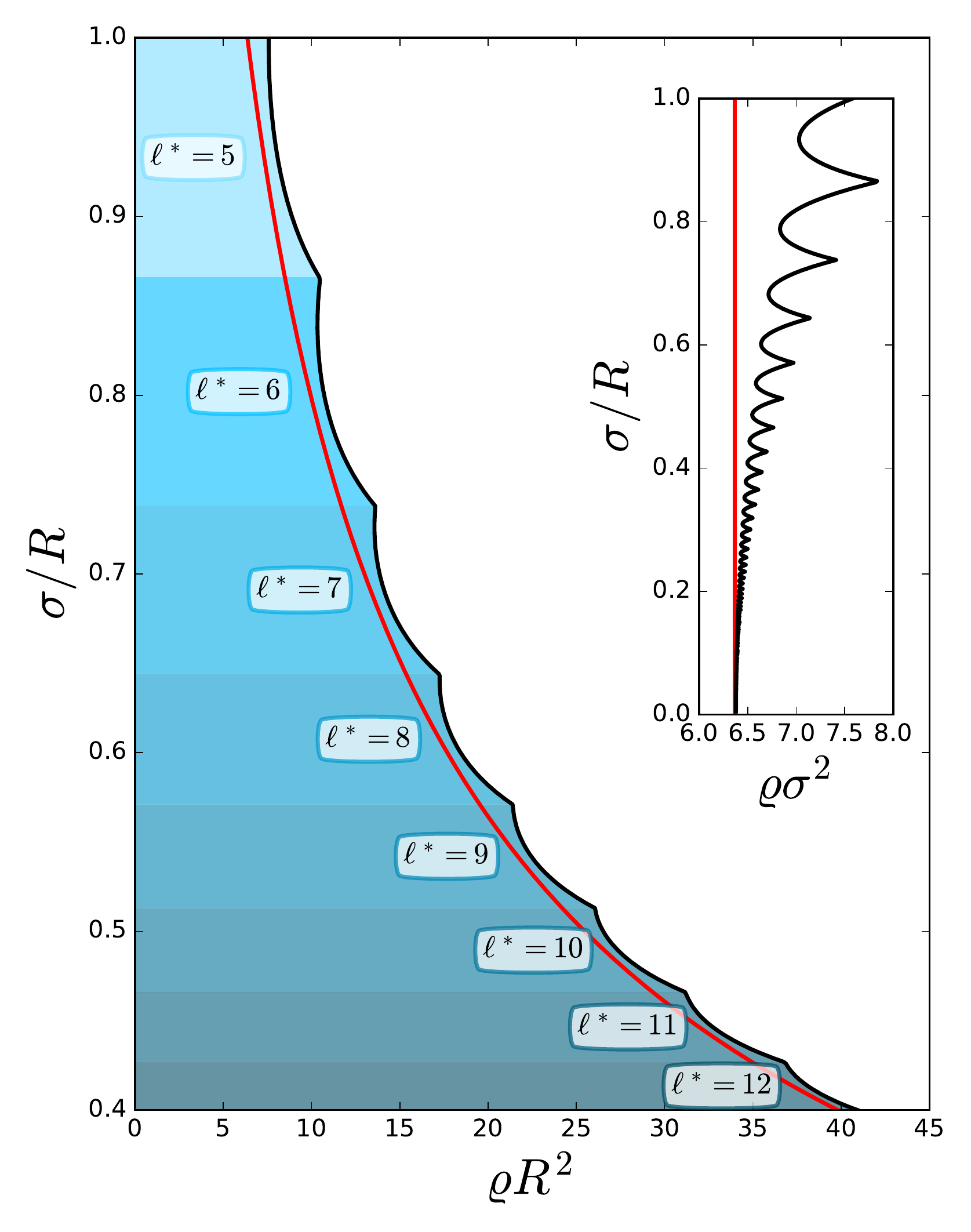}
\caption{Plot of the $\lambda$-line at fixed temperature $T^*=1$ of the spherical system (black line). At fixed radius $R$, the $\lambda$-line can be compared to that of a planar system with the same range of the potential $\sigma$ (red line). In the inset the density is rescaled in order to better highlight the differences between the planar and spherical systems. The peculiar kinks appearing in the $\lambda$-line of the spherical system are due to the changing position $\ell^*$ of the negative minimum in the energy spectrum of the potential; the different shades of blue in the background highlight the regions where $\ell^*$ is constant.}
\label{lambda_gem4}
\end{figure}

\section{SIMULATION}
\label{simulation}

To test the validity of the theoretical results obtained through DFT, we performed Monte Carlo (MC) simulations of the GEM-4 fluid in the canonical ensemble, that is, at fixed number of particles $N$, surface area $A$, and temperature $T$. Notice that, since the system we are considering is rigorously finite in size, to have a complete picture of the behavior of the fluid we must control separately the surface area $A$ (given by $4\pi R^2$ where $R$ is the sphere radius) and the number of particles $N$, rather than just fixing the number of particles and controlling the density of the system by varying the surface area.

As a consequence of this, most of the states sampled through our MC simulations require only a small number of particles ($N \leq 500$), and reliable results can be obtained even with short simulations. Still, to speed up the onerous computations of the distances between the particles and their interaction potential, we use lookup tables computed once at the beginning of the simulation, so that evaluation of the distance $r$ between two particles only requires the computation of $\x\cdot\x' = \cos ( r/R )$.

The moves of the MC are simple rotations of a particle around its position. More specifically, we consider a particle with coordinates $\x = [\cos\phi \sin\theta, \sin\phi \sin\theta, \cos\theta]$ and use its position as the north pole of a new coordinate system, so that its new coordinates are $\x_0 = [0, 0, 1]$. We then randomly choose a new value $\cos \theta'$ for its third component $z$ in $[1-\delta,1]$, where the parameter $\delta$ sets the maximum stride of the move, and a direction for the move $\phi' \in [0,2\pi]$, so that its new coordinates are $\x'_0 = [ \cos\phi' \sin\theta', \sin\phi' \sin\theta', \cos\theta']$. Finally we return to the previous coordinate system by applying the rotation that gives $\hat{R}\x_0 = \x$ so that we obtain the new coordinates of the particle as $\x' = \hat{R}\x'_0$.

\begin{figure}[t]
\includegraphics[width = 3in]{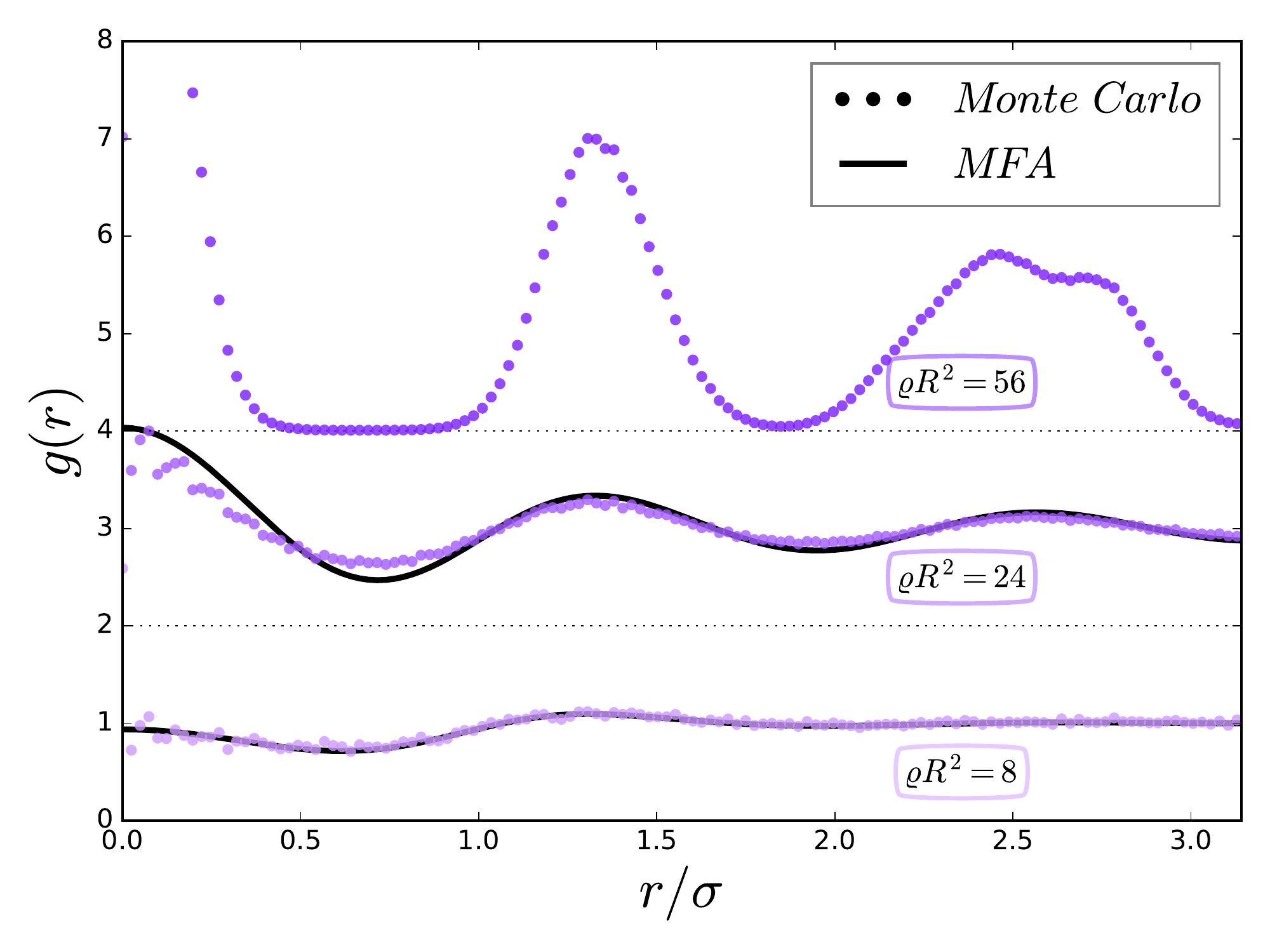}
\caption{Comparison between the theoretical and simulated two-point correlation functions at $\sigma/R = 0.5$ and $T^* = 1$ for increasing densities $\varrho$. The correlation functions at $\varrho R^2 = 8$ and $\varrho R^2 = 24$ refer to the fluid phase, while that at $\varrho R^2 = 56$, showing only the MC simulation data, refers to a crystallized system. The plots were offset for clarity; the dotted line shows the offset x axes.}
\label{corr_gem4}
\end{figure}

\section{RESULTS}\label{THREE}

\subsection{The homogeneous fluid}
\label{fluid}

We start our discussion of the results by looking at the properties of the low-density homogeneous fluid.

To delimit the portion of the phase diagram we are interested in, we first draw the $\lambda$-line given by equation \eqref{lambda} at fixed reduced temperature $T^*=1$. This is the boundary beyond which the homogeneous solution of the Euler-Lagrange equation is no longer a local minimum of the mean field grand potential $\beta\Omega$, meaning that this fluid state cannot be an equilibrium state.

In figure \ref{lambda_gem4} we show the result of this operation. By fixing the radius $R$ of the sphere and varying the potential range $\sigma$, we can compare the $\lambda$-line obtained for the spherical system to the one for bulk planar fluids. In the planar case, $\sigma$ is the only length scale of the system, so the density of the $\lambda$-line is simply proportional to $1/\sigma^2$.

On the contrary, in the spherical case, the presence of two length scales gives origin to a more complex $\lambda$-line with a saw-like structure.
Since the potential varies continuously as the ratio $\sigma/R$ which controls its behavior is changed, it is difficult to understand the provenance of the kinks in the $\lambda$-line. However, the explanation is plain once one considers the spectrum $w_{\ell}$ of the potential.

In fact, one finds that the kinks correspond to the values of $\sigma/R$ at which the negative minimum of $w_{\ell}$ shifts position. As we will see, the position $\ell^*$ of the minimum affects not only the properties of the homogeneous fluid, but also those of the high-density cluster crystal phases.

The inset in figure \ref{lambda_gem4} also shows that as $\sigma/R \rightarrow 0$, the $\lambda$-lines for the planar and spherical fluid converge, as is expected from the fact that a sphere of infinite radius approximates the plane. From this point of view, we can take the ratio $\sigma/R$ as a measure of how much the curvature of the sphere affects the fluid behavior.

For the low-density region delimited by the $\lambda$-line, we can obtain theoretical correlation functions from equation \eqref{OrnZern} in the mean field approximation \eqref{mfa}, and compare them with those obtained from MC simulations. In figure \ref{corr_gem4} we show the correlation function $g(r)=h(r)+1$ for two densities at which the system is fluid, and the correlation function obtained from a simulation in the high-density regime.

We see that the theoretical predictions closely resemble the correlation functions obtained from MC simulations. The discrepancies grow as the density of the fluid approaches the $\lambda$-line, but the theoretical $g(r)$ still correctly predicts the formation of clusters even in the homogeneous phase, as can be seen from the peak of the correlation function at $r=0$. This is a known feature of the system also for the bulk fluid \cite{Mladek_2007}. We also observe in figure \ref{corr_gem4} the appearance of fluctuations with a definite length scale, which become the clusters in the high density, inhomogeneous phase. 

\begin{figure}[t]
\includegraphics[width = 3in]{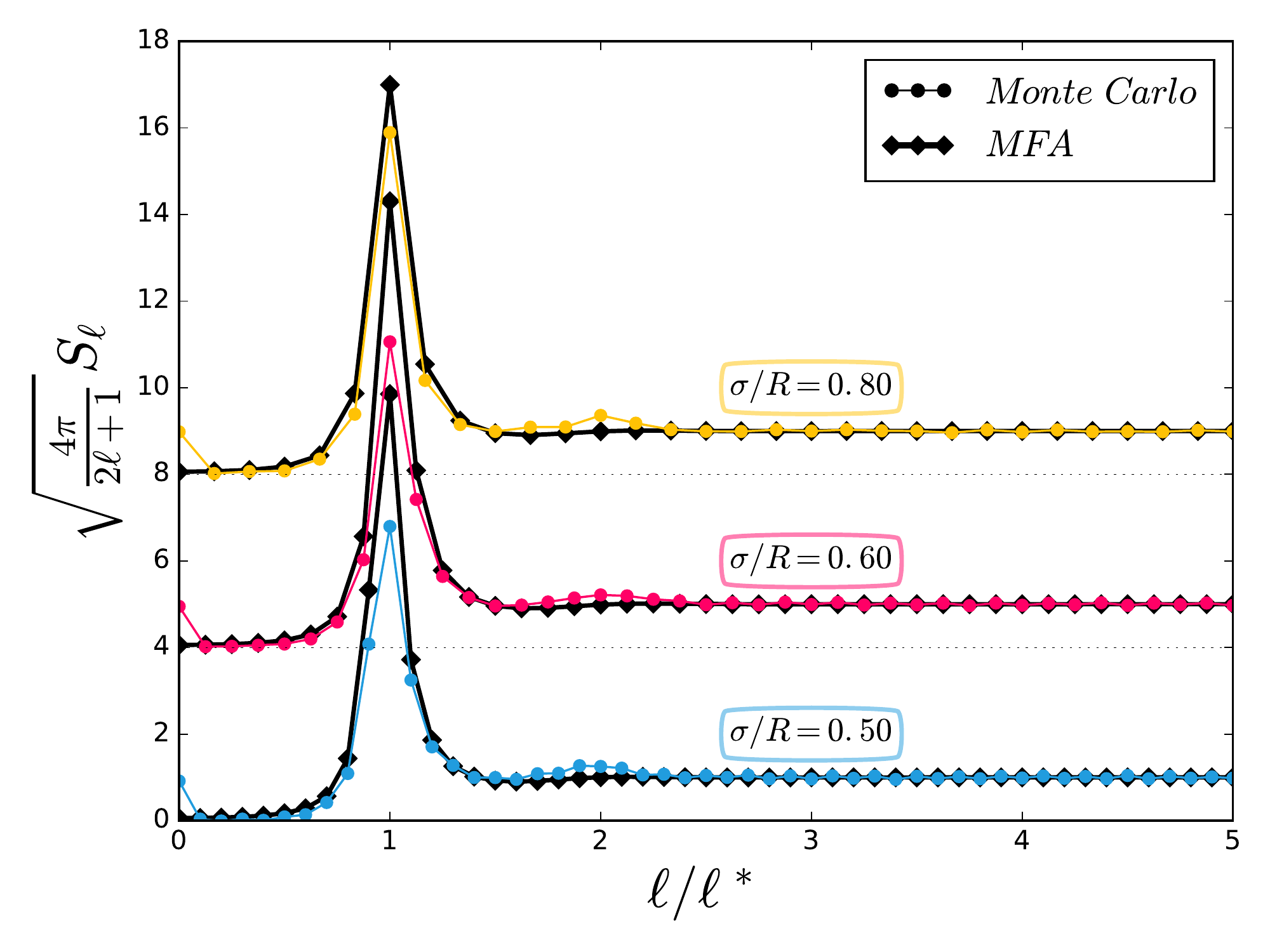}
\caption{Structure factor at different $\sigma/R$ at $T^* = 1$. The mean density of the homogeneous fluid is fixed at $\varrho \sim 0.95 \varrho_{\lambda}$. The plots were offset for clarity, the dotted line shows the offset of the x axes.}
\label{str_gem4}
\end{figure}

A qualitative explanation of this behavior is that the energetic cost a particle pays for overlapping completely with a few other particles inside a cluster is less than the cost of partially overlapping with many particles \cite{Mladek_2008_bis}. As the density becomes larger, it becomes convenient to form clusters separated by energy barriers resulting from the cumulative repulsions of multiple clusters.

Notice that in the high-density phase the interstitial space between two consecutive peaks of the $g(r)$ is depleted of particles, so that clusters can be considered as clear-cut ``objects'' with a sharp interface. By contrast, in the homogeneous fluid, clusters must be regarded as polydisperse, dynamical entities in chemical equilibrium with isolated particles.

The discussion about the length scales of the oscillations of the correlation function can be made more quantitative from inspection of the structure factors, which we plot for different values of $\sigma/R$ in figure \ref{str_gem4} at densities close to the $\lambda$-line, namely at $\varrho \sim 0.95 \varrho_{\lambda}$.

For densities so close to the $\lambda$-line, there are some discrepancies between theory and simulation, in particular at $\ell=0$ and $\ell=\ell^{*}$, 
where the theory respectively underestimates and overestimates $S_{\ell}$, as well as at $\ell \sim 2\ell^*$, where the secondary bump of $S_{\ell}$ obtained from the simulation is not reproduced by the theory. Nevertheless we observe that the main expected feature of the structure factor, namely the presence of a peak at $\ell=\ell^*$, is present in both theory and simulation.

The presence of this single peak in correspondence to the position of the potential spectrum minimum $\ell^*$, shows that there is a single prominent spherical harmonic component of the correlation function $g(r)$, originating fluctuations with characteristic length scale $2\pi R/\ell^*$ which also set the expectation value of the inter-cluster distance in the cluster crystal. We also recall that the Hansen-Verlet freezing criterion states that when the height of the principal peak of the structure factor is larger than the threshold value $\sim 2.8$, the system is frozen \cite{March_2002}. A characteristic of the GEM-4 fluid is that this peak becomes much higher than that observed for fluids with hard repulsions, showing that soft repulsions can sustain a larger degree of spatial correlation before actually freezing \cite{Mladek_2007}.

\begin{figure}[t]
\includegraphics[width = 3in]{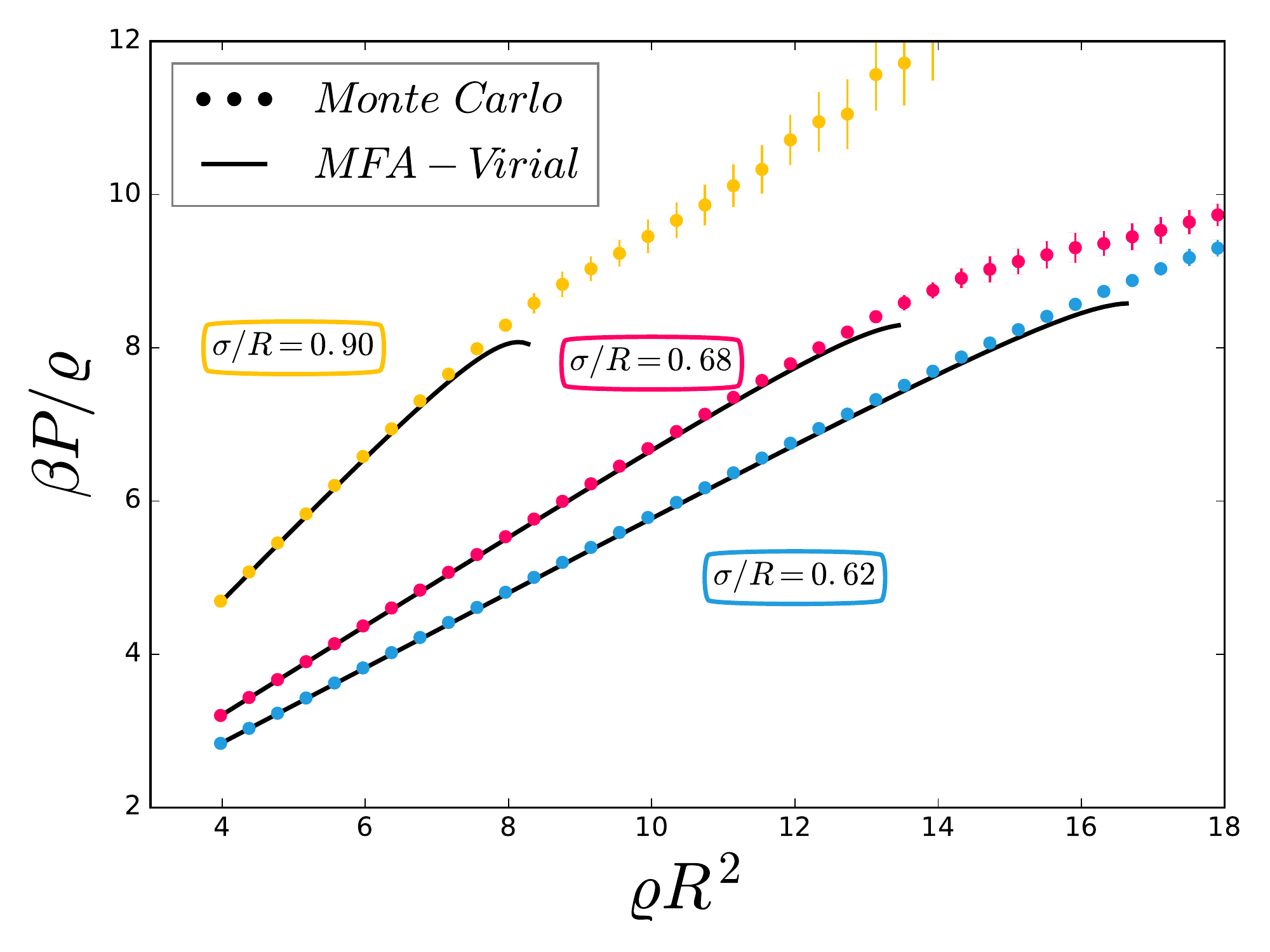}
\caption{Compressibility factor for different values of $\sigma/R$. Theoretical predictions for the homogeneous fluid obtained from the correlation function via the virial route are compared to data from MC simulations. The reduced temperature is fixed at $T^*=1$.}
\label{gem4_eqstate}
\end{figure}

We complete our review of the properties of the fluid state by studying the equation of state in the low density regime.

To compute the pressure in the MC simulation, we use the usual virial method. However, in the case of spherical systems the definition of the virial depends on the convention used to compute distances \cite{Post_1986}. Having used the curved line of force convention, we have

\begin{equation}
	P = \frac{\varrho}{\beta} - \frac{1}{8\pi} < \sum_{i\neq j}  w'(\x_i\cdot\x_j)\arccos ( \x_i\cdot\x_j) >
\end{equation}

where $w'(\x_i\cdot\x_j)$ is the derivative of the potential and the angled brackets denote ensemble average.

In figure \ref{gem4_eqstate} we compare the results for the compressibility factor $\beta P/\varrho$ with the theoretical predictions obtained from the correlation function $g(r)$ through the virial route. The simulation results are closely matched at low density, but the agreement rapidly deteriorates in proximity of the $\lambda$-line, where the theory encounters a singularity while the fluid in the simulated system freezes to become the cluster crystal phase expected at high densities, as testified by the change in the slope of the compressibility factor which signals the transition.

\begin{figure}[t!]
\centering
\includegraphics[width = 3in]{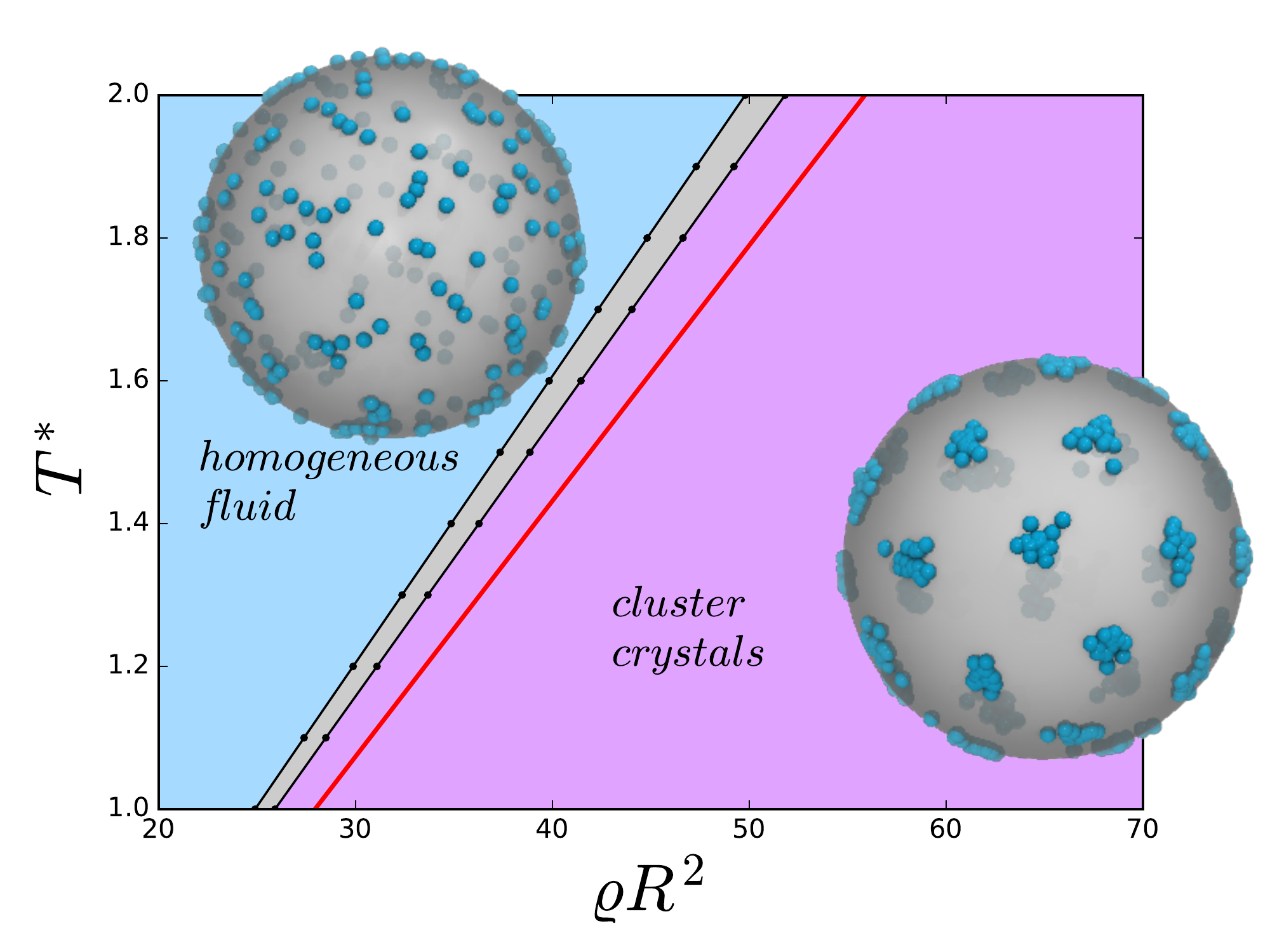}
\caption{Theoretical predictions for the phase diagram of the GEM-4 fluid in the $\varrho$-$T^*$ plane, at fixed $\sigma/R = 0.5$. The solid red line is the $\lambda$-line. The filled points are transition boundaries, solid black lines are a guide for the eye. As the mean density increases, the homogeneous phase on the left is replaced by a cluster crystal configuration with 32 clusters. The gray area is the transition region. Snapshots of the MC simulations are added to illustrate the two phases, with the centers of the particles represented as blue beads on the surface of a sphere. Notice that the radius of the beads in the snapshots was chosen to enhance visibility and has no physical meaning.} 
\label{diag_2_gem4}
\end{figure}

\begin{figure}[t!]
\includegraphics[width = 3in]{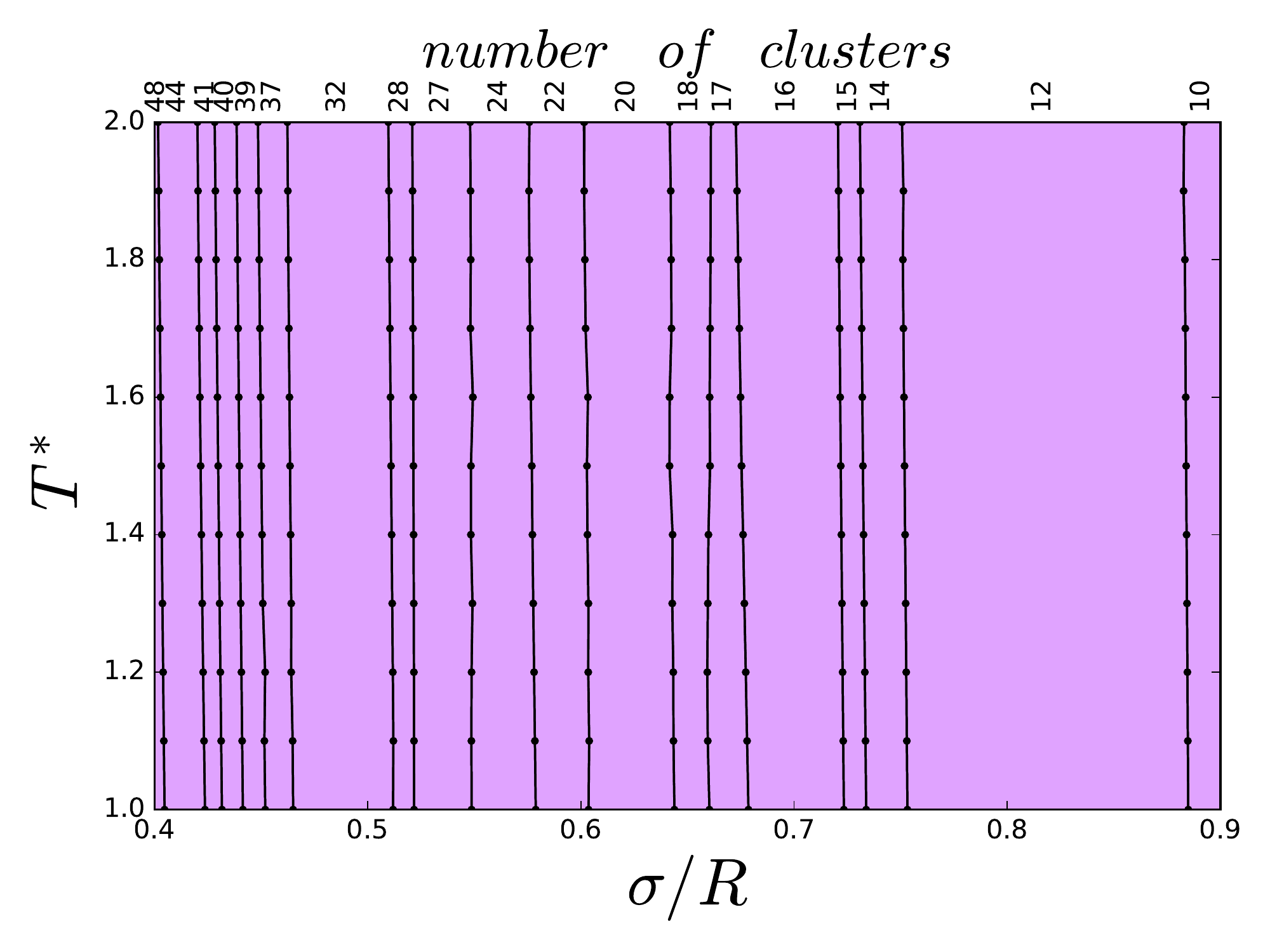}
\caption{Theoretical predictions for the phase diagram of the GEM-4 fluid in the $\sigma/R$-$T^*$ plane, at fixed $\varrho R^{2}=87.0$. The filled points are transition boundaries, solid lines are a guide for the eye. Different cluster phases are encountered as $\sigma/R$ increases, and the number of clusters of each phase is indicated in the upper x-scale.}
\label{diag_1_gem4}
\end{figure}

\begin{figure}[t!]
\includegraphics[width = 3in]{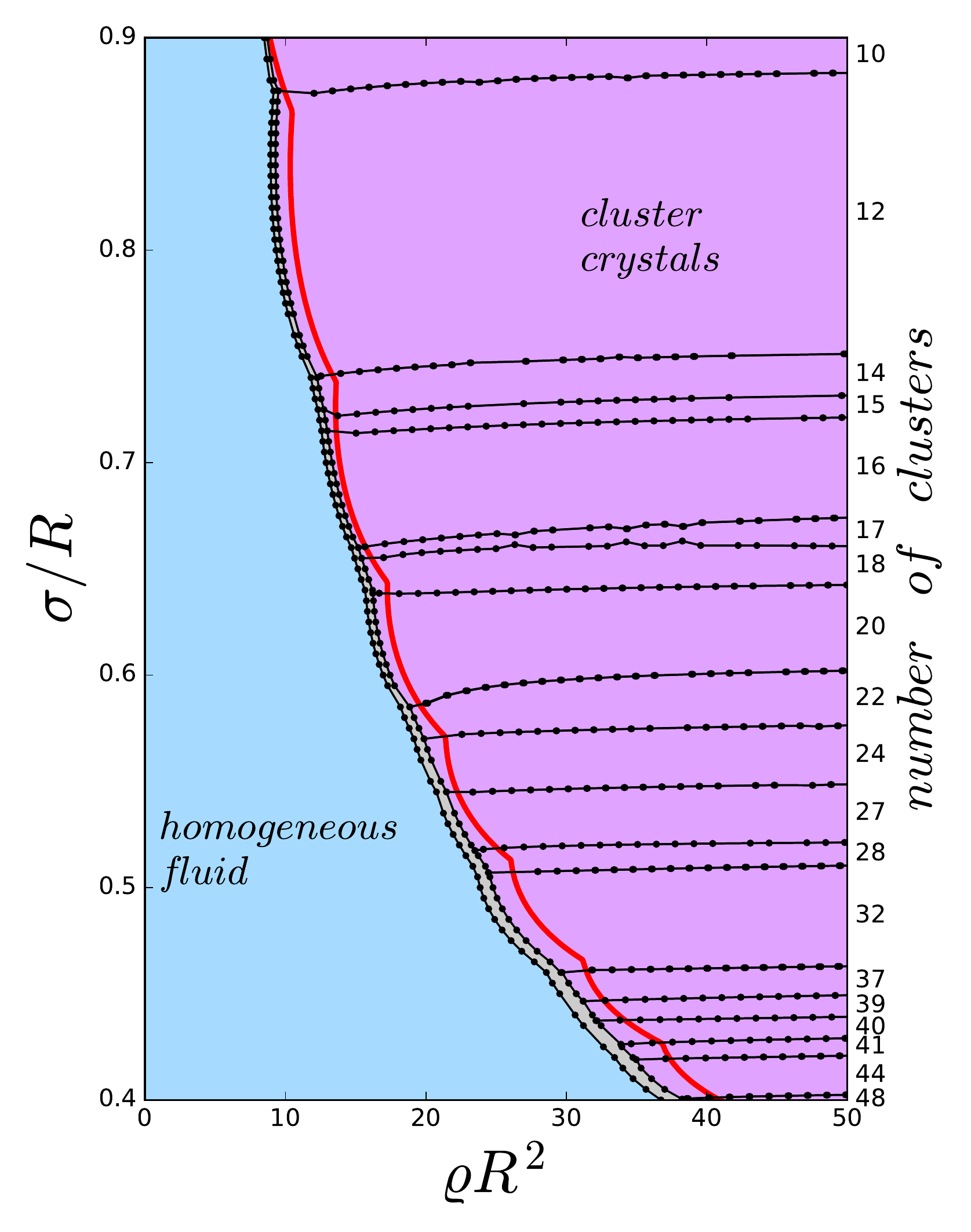}
\caption{Theoretical predictions for the phase diagram of the GEM-4 fluid in the $\varrho$-$\sigma/R$ plane, at fixed $T^*=1$. The solid red line is the $\lambda$-line, filled points are the first order phase transition boundaries, and solid black lines are a guide for the eye. The gray area is the transition region. At low densities the homogeneous phase (pink) is stable, while at greater densities it is replaced by the more stable cluster crystal phases. Different cluster phases are encountered as $\sigma/R$ is changed, the number of clusters of each phase is indicated in the right y-scale.}
\label{diag_0_gem4}
\end{figure}

\subsection{The phase diagram}
\label{phase}

We obtain theoretical predictions for the phase diagram of the GEM-4 fluid based on the DFT results. We study the behavior of the system by changing the mean density $\varrho$, the ratio $\sigma/R$ and the temperature $T^*$. Since the functional minimization was carried out in the grand canonical ensemble, the mean density $\varrho$ was obtained \textit{a posteriori} as a function of the chemical potential $\mu$. As discussed in section \ref{DFT}, for each value of $\mu$ the minimization was carried out starting from multiple trial density profiles, resulting in multiple output distributions, from which we selected the most stable one. 

\begin{figure}[t]
\centering
\includegraphics[width = 3in]{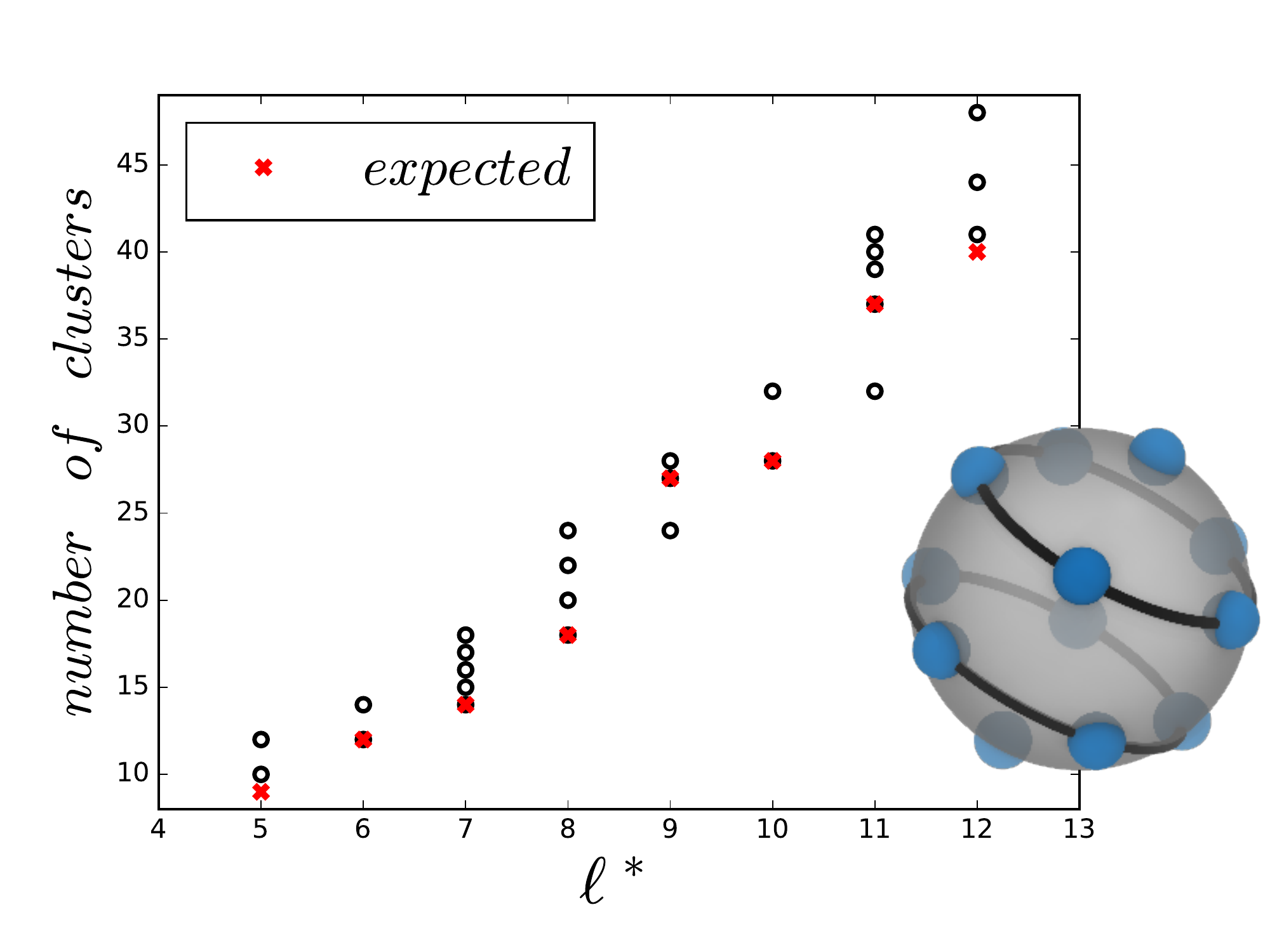}
\caption{Number of clusters as a function of the position $\ell^*$ of the energy spectrum minimum. The expected values computed using our shell model are plotted along with the observed stable cluster configurations. The inset image shows an example of the shell method employed to predict the number of clusters from $\ell^*$, with the predicted clusters displayed as blue beads arranged on the black shells.}
\label{clus_est}
\end{figure}

In figures \ref{diag_2_gem4}, \ref{diag_1_gem4}, and \ref{diag_0_gem4} we show respectively the phase diagram at fixed range ($\sigma/R = 0.5$), at fixed mean density ($\varrho = 87.0$), and at fixed temperature ($T^* = 1.0$). Figure \ref{diag_2_gem4} also shows the snapshots of MC simulations of the system in the low density homogeneous state and in the high density cluster crystal configuration.

The boundaries between two phases $A$ and $B$ are obtained by a Maxwell construction, i.e., by imposing the conditions $\mu_A = \mu_B$, $\beta\Omega_A = \beta\Omega_B$. However, we stress the fact that in a finite system, such as a fluid on the surface of a sphere, we cannot have real phase transitions, and even metastable states can contribute to the thermodynamics of the system. In particular, the melting and freezing lines which we obtain through the Maxwell construction, do not delimit a \textit{bona fide} coexistence region: in fact, since we are considering a finite system, the energy penalty for creating interfaces between different phases prevents the formation of a phase separated system. Rather, we can regard this region as an approximation of the crossover region, in which the system displays intermediate properties between the fluid and the crystal phases, along the lines developed in ref. \cite{Hill_1962}. In the thermodynamic limit (the planar system) the two regions rigorously coincide \cite{Hill_1962}. Similar considerations hold in principle also for the transition between different crystal phases, although in this case the transition region is extremely narrow, and hardly distinguishable from the line along which the Helmholtz free energies of the phases at hand coincide at the same $\rho$.

In the $\varrho - T^*$ plane, the behavior of the system can be compared to the bulk case. We see that not only the $\lambda$-line, but also the freezing line is linear in this plane to a high degree. This is a known characteristic of the GEM-4 fluid \cite{Mladek_2007, Prestipino_2014}, and an analytical justification of this behavior is given in reference \cite{Kahl_2007}.

An aspect which can only be studied in spherical systems is the competition between different cluster crystal configurations, which we label by the number of clusters they form. As seen in figures \ref{diag_1_gem4} and \ref{diag_0_gem4}, the number of clusters of the most stable configuration is nearly independent of the temperature or mean density of the system, especially away from the freezing transition. Thus the only parameter that controls the number of clusters formed by the fluid on the sphere surface is the ratio $\sigma/R$.

Moreover, we observe that some cluster configurations span larger intervals of the $\sigma/R$ axis, and transitions roughly correspond to the points where the energy spectrum minimum position changes.

With this observation in mind, we ask whether it is possible to make a rough prediction of the number of clusters using only the information about $\ell^*$. To do so, we assume that clusters are arranged on shells going from one pole to the other and corresponding to the maxima of the zonal spherical harmonic $Y_{\ell^*}(\theta)$, which roughly correspond to the maxima of the correlation function.

If clusters are $d \simeq 2\pi R/\ell^*$ away from each other, then each shell can host a number of clusters equal to

\begin{equation}
	n_{shell} \simeq \frac{2\pi R \sin(\theta_{shell})}{2\pi R/\ell^*} = \ell^* \sin(\theta_{shell}),
\end{equation}

where $\theta_{shell}$ is the position of the local maximum in the zonal harmonic $Y_{\ell^*}(\theta)$ corresponding to the shell under consideration. The total number of clusters is simply given by the sum over the shells.

In figure \ref{clus_est} we plot the number of clusters as a function of the position $\ell^*$ of the energy spectrum minimum, and compare it with our rough prediction. Even with this simple method we are able to give a good estimate of the number of clusters found for a given range of the potential, although it cannot distinguish between different configurations found at the same $\ell^*$.

We do not provide simulation results for the transition lines and the most stable cluster configurations, because their precise computation requires more advanced techniques than simple MC simulations, such as Widom insertion method \cite{Widom_1963} or thermodynamic integration, which has been used before in the same context \cite{Mladek_2007_bis}. This kind of investigation goes beyond the scope of the present paper. 

Nevertheless, we remark that our MC results are consistent with the theoretical phase diagrams presented in this section. In fact, starting from a homogeneous configuration at densities above the theoretical freezing line, one observes the spontaneous formation of cluster crystals; on the other hand, starting from cluster crystals, they spontaneously melt at densities below the theoretical freezing line. This is revealed both by visual inspection of the simulation trajectories and by inspection of the correlation function $g(r)$.

\begin{figure*}
 \centering
 \subfloat[10 clusters]{\includegraphics[width = 2.7cm]{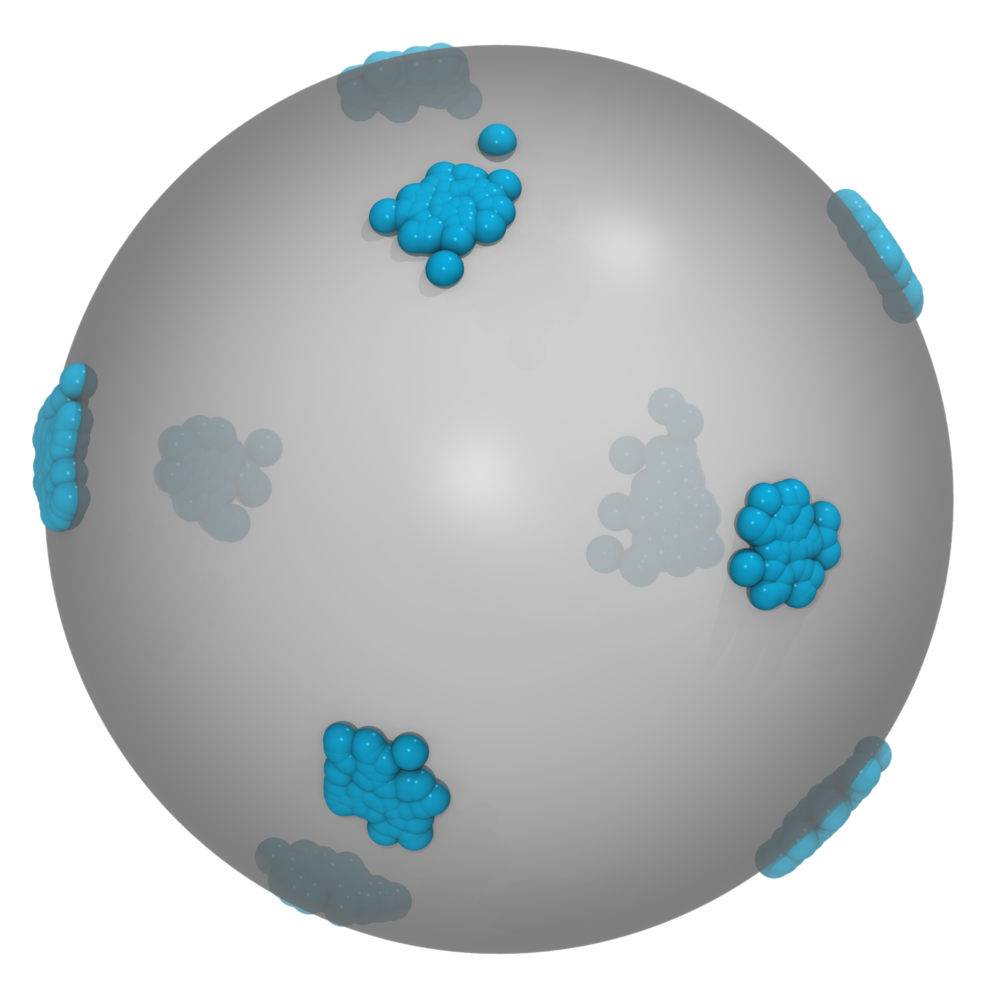}}
 \includegraphics[width = 2.7cm]{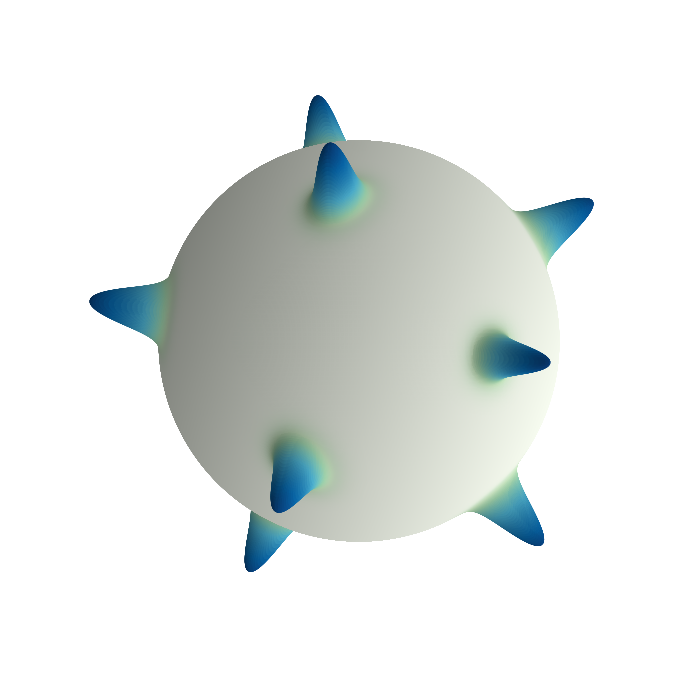}
 \includegraphics[width = 2.7cm]{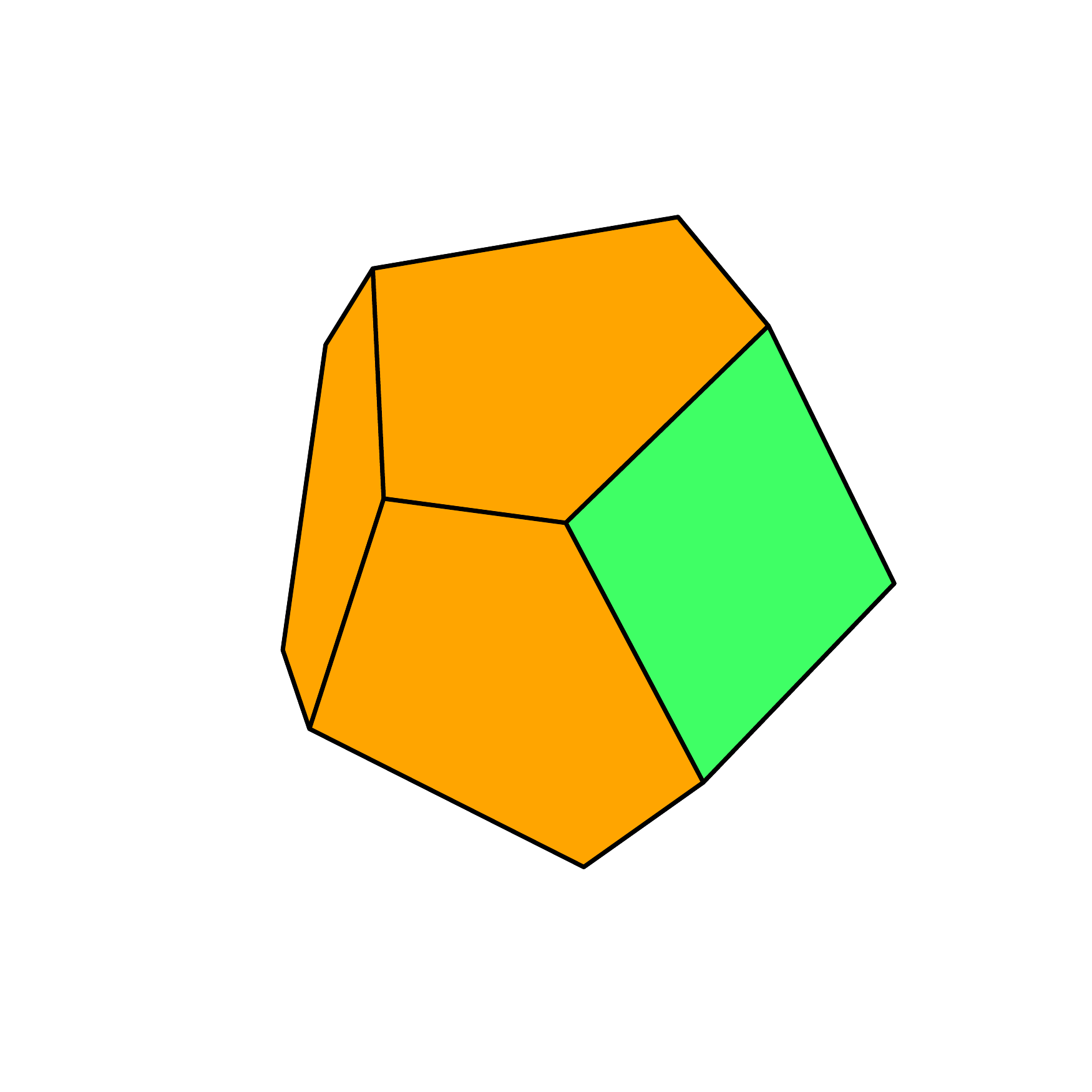}
 \subfloat[16 clusters]{\includegraphics[width = 2.7cm]{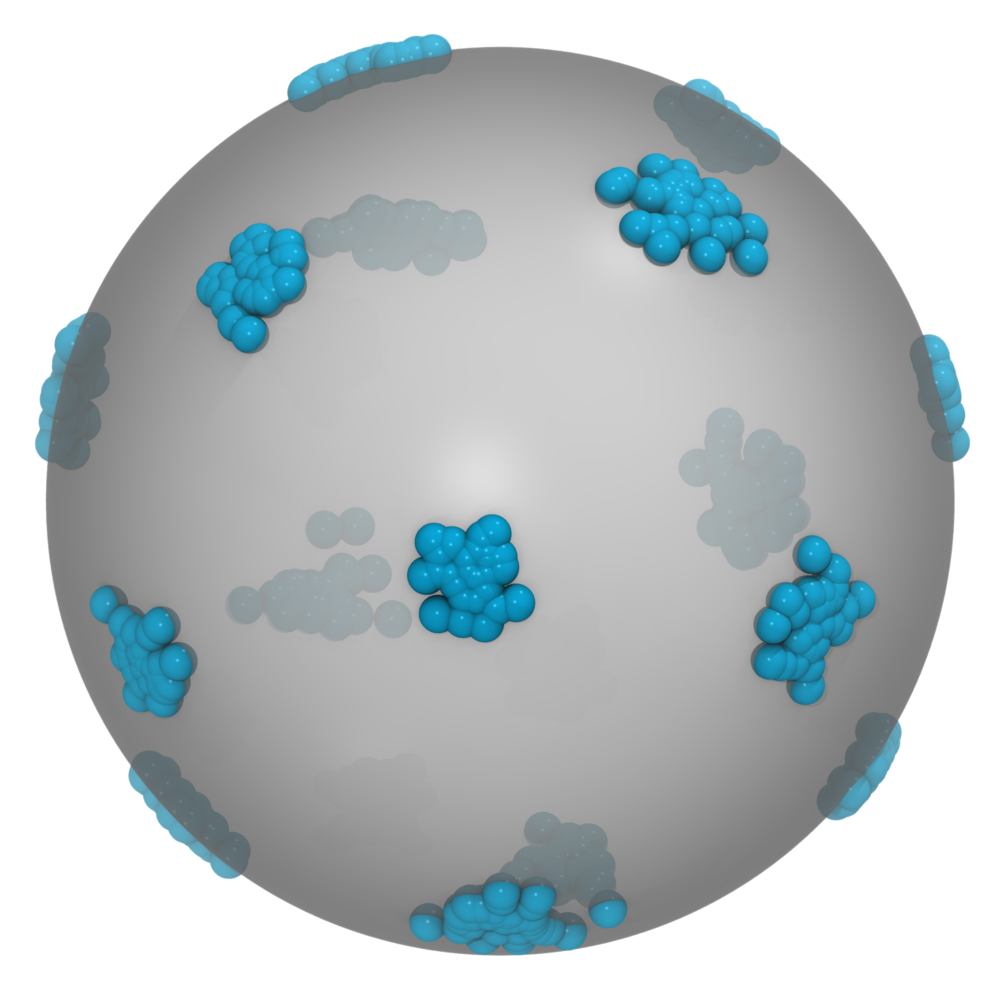}}
 \includegraphics[width = 2.7cm]{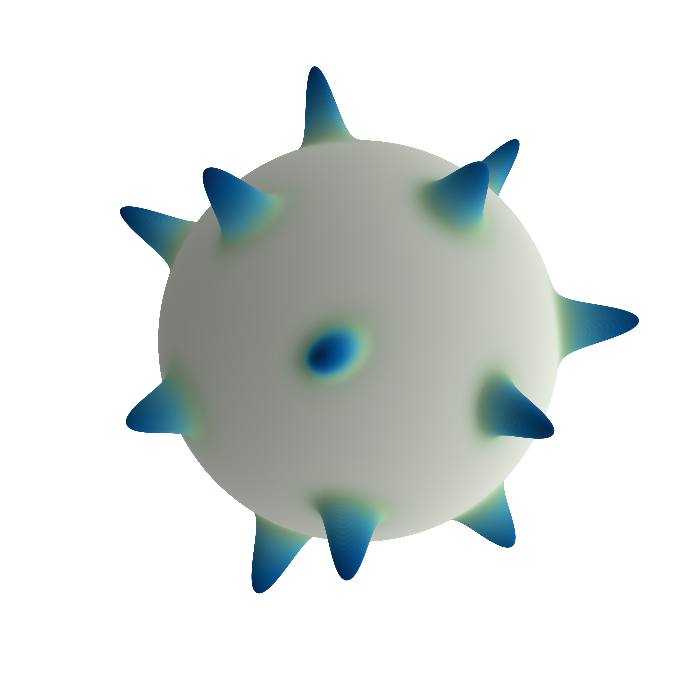}
 \includegraphics[width = 2.7cm]{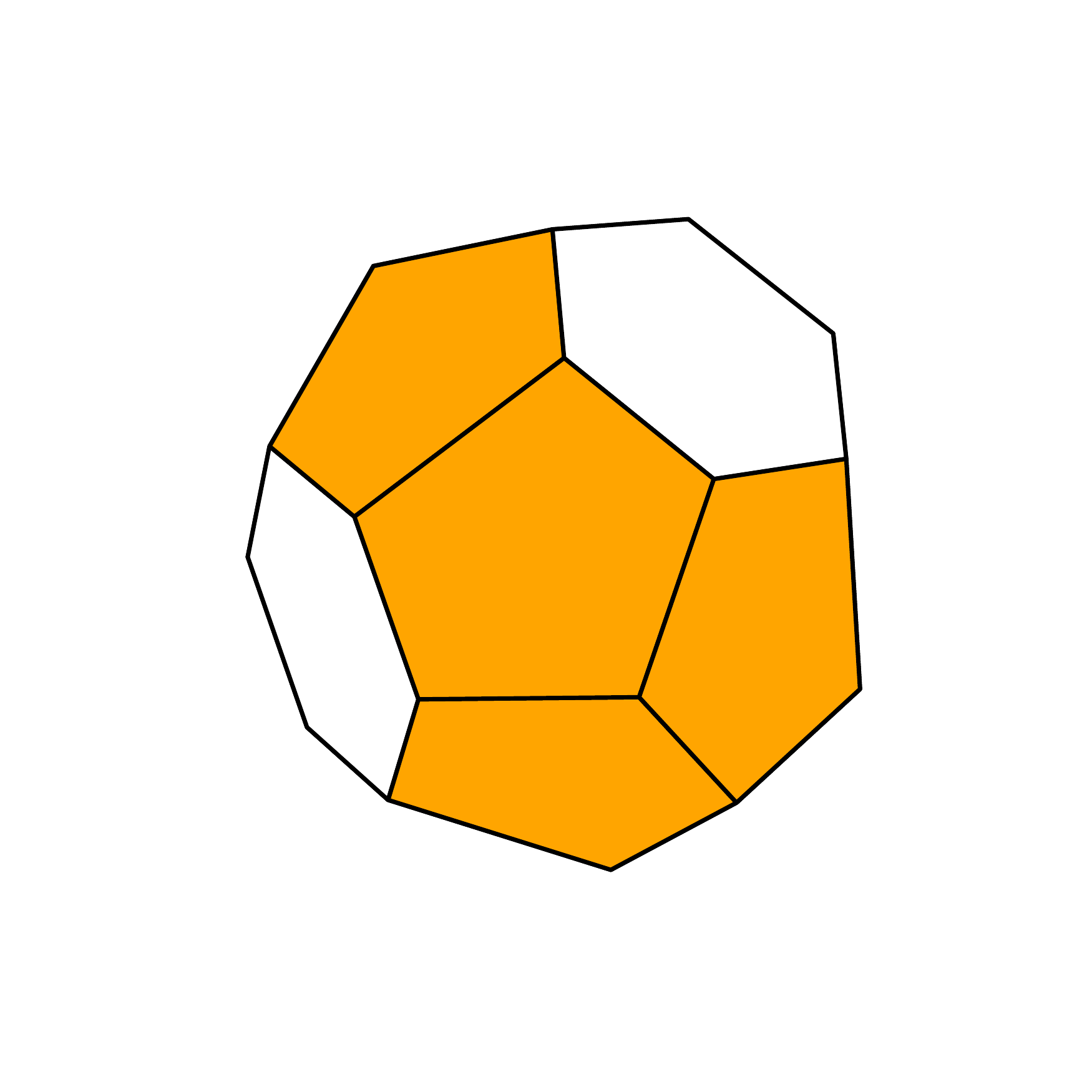}
 \includegraphics[width = 1.5cm]{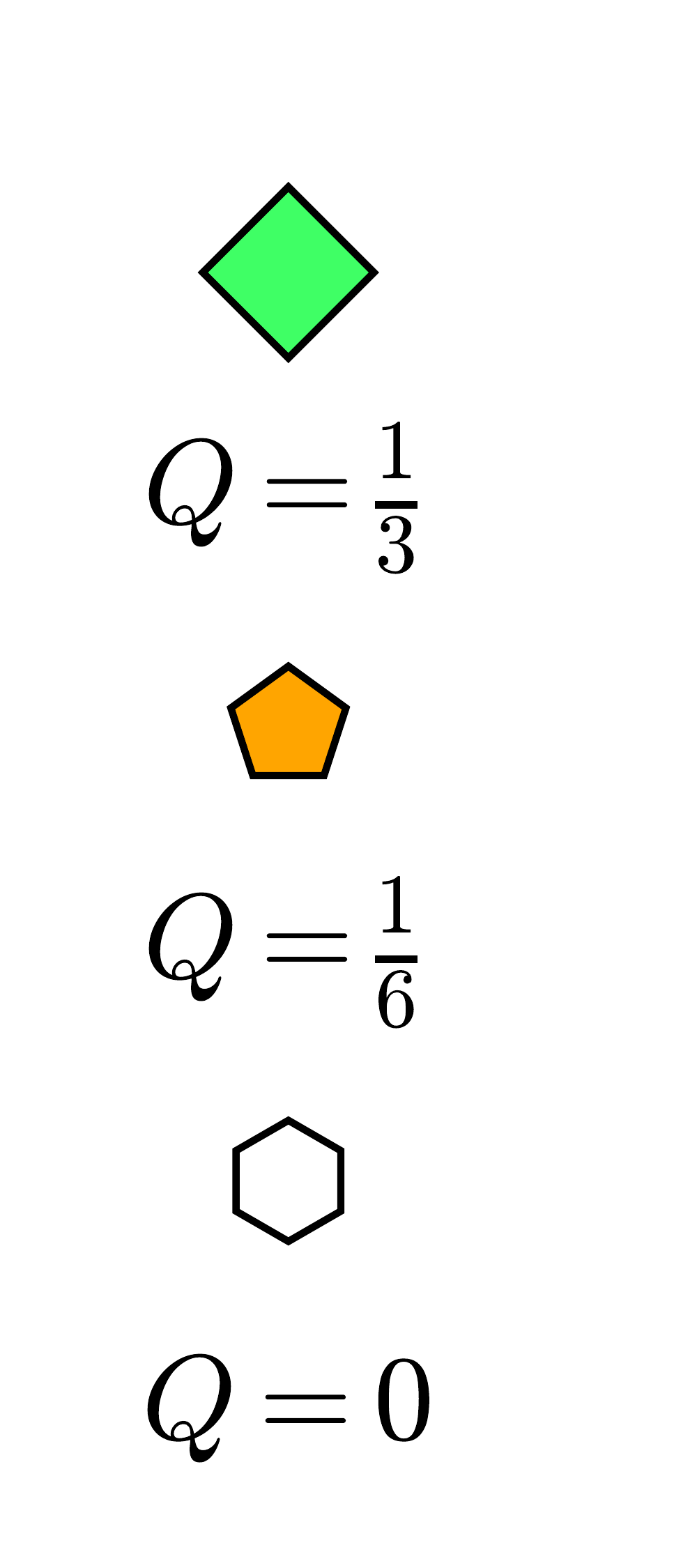}\\
 \subfloat[20 clusters]{\includegraphics[width = 2.7cm]{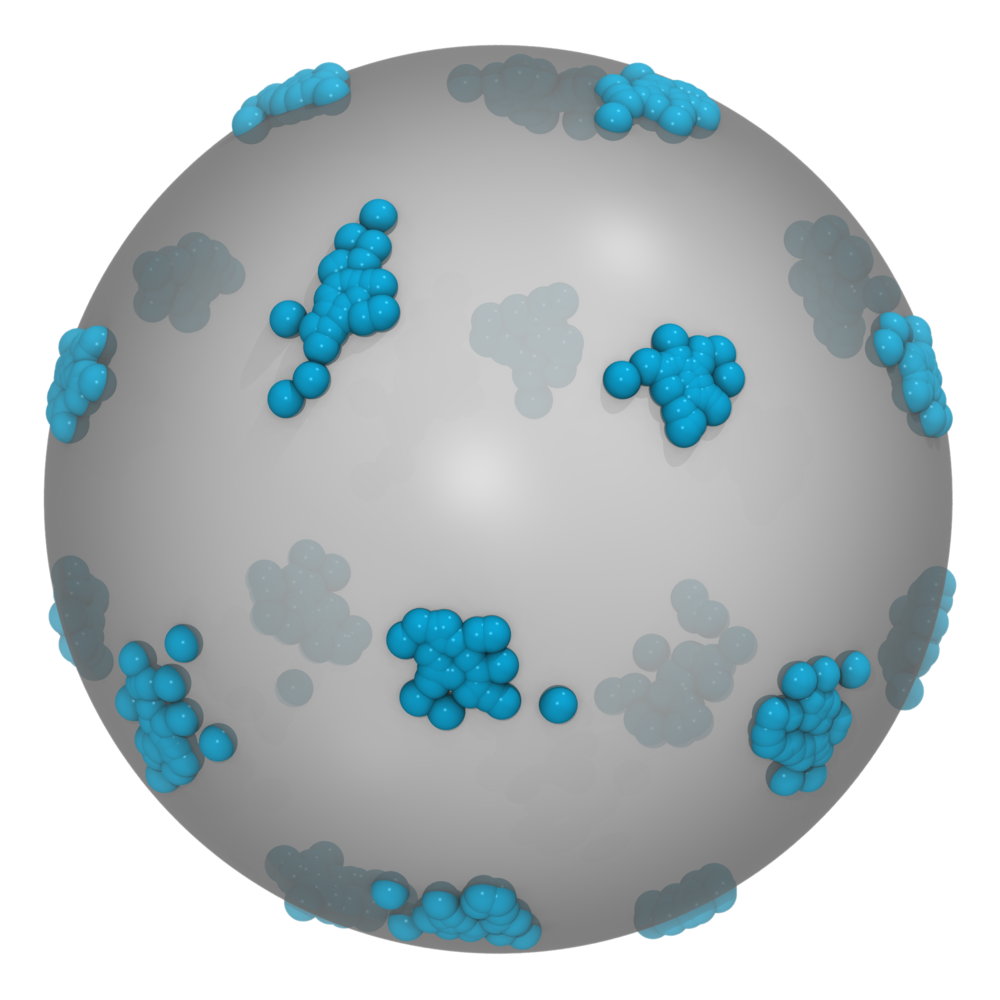}}
 \includegraphics[width = 2.7cm]{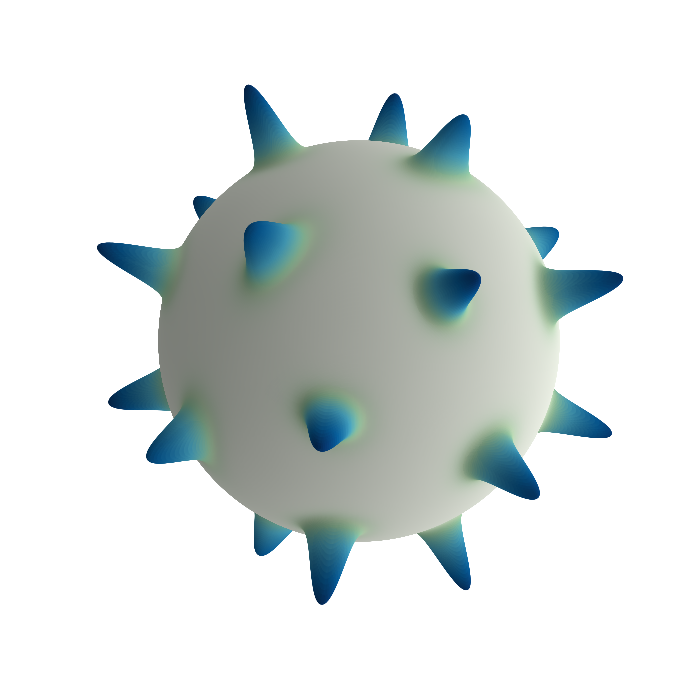}
 \includegraphics[width = 2.7cm]{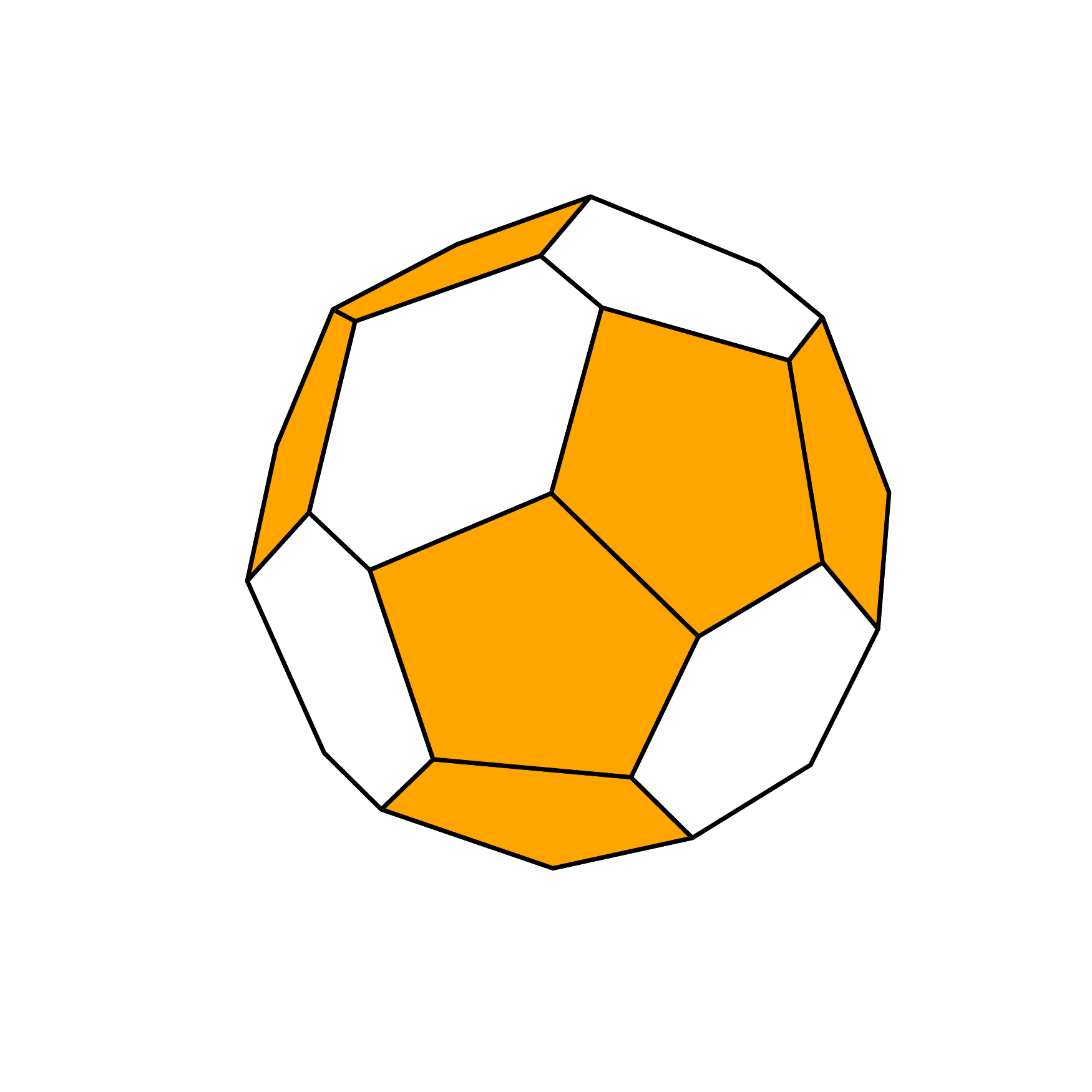}
 \subfloat[32 clusters]{\includegraphics[width = 2.7cm]{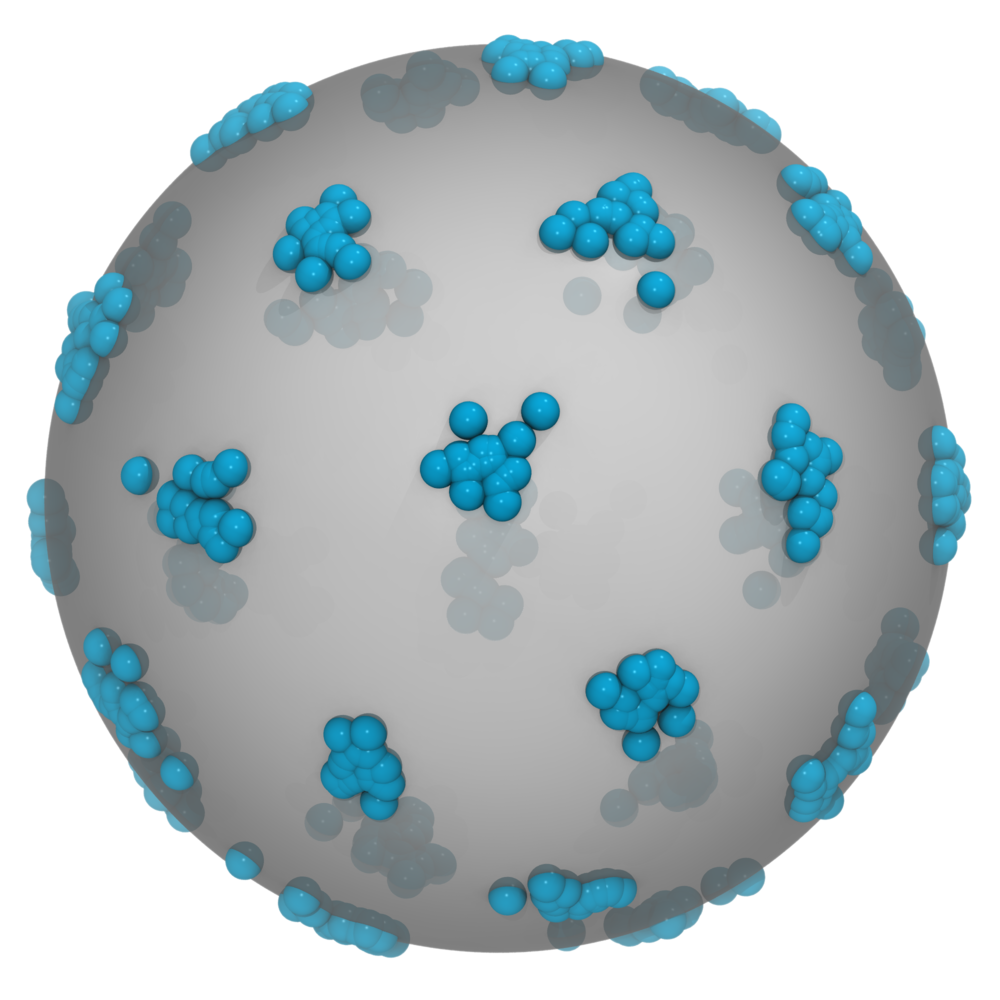}}
 \includegraphics[width = 2.7cm]{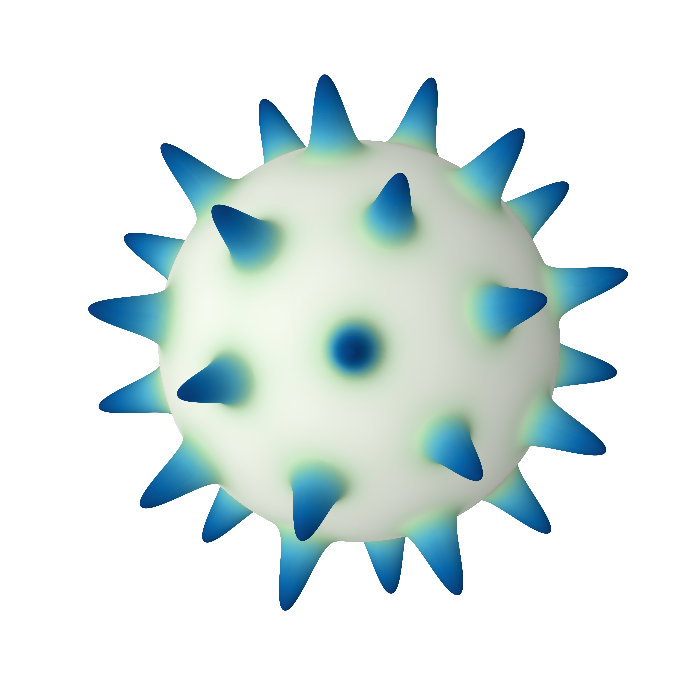}
 \includegraphics[width = 2.7cm]{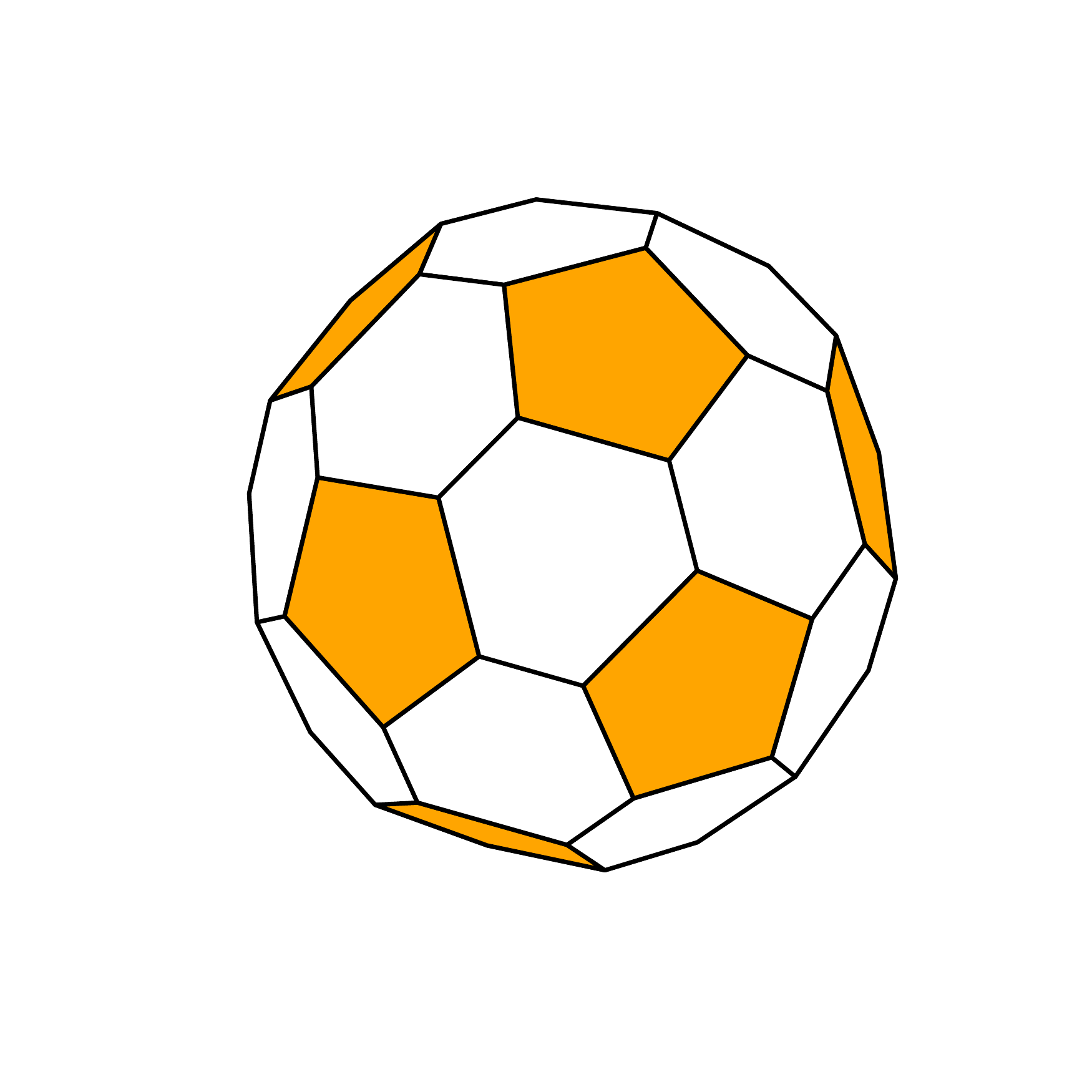}
 \includegraphics[width = 1.5cm]{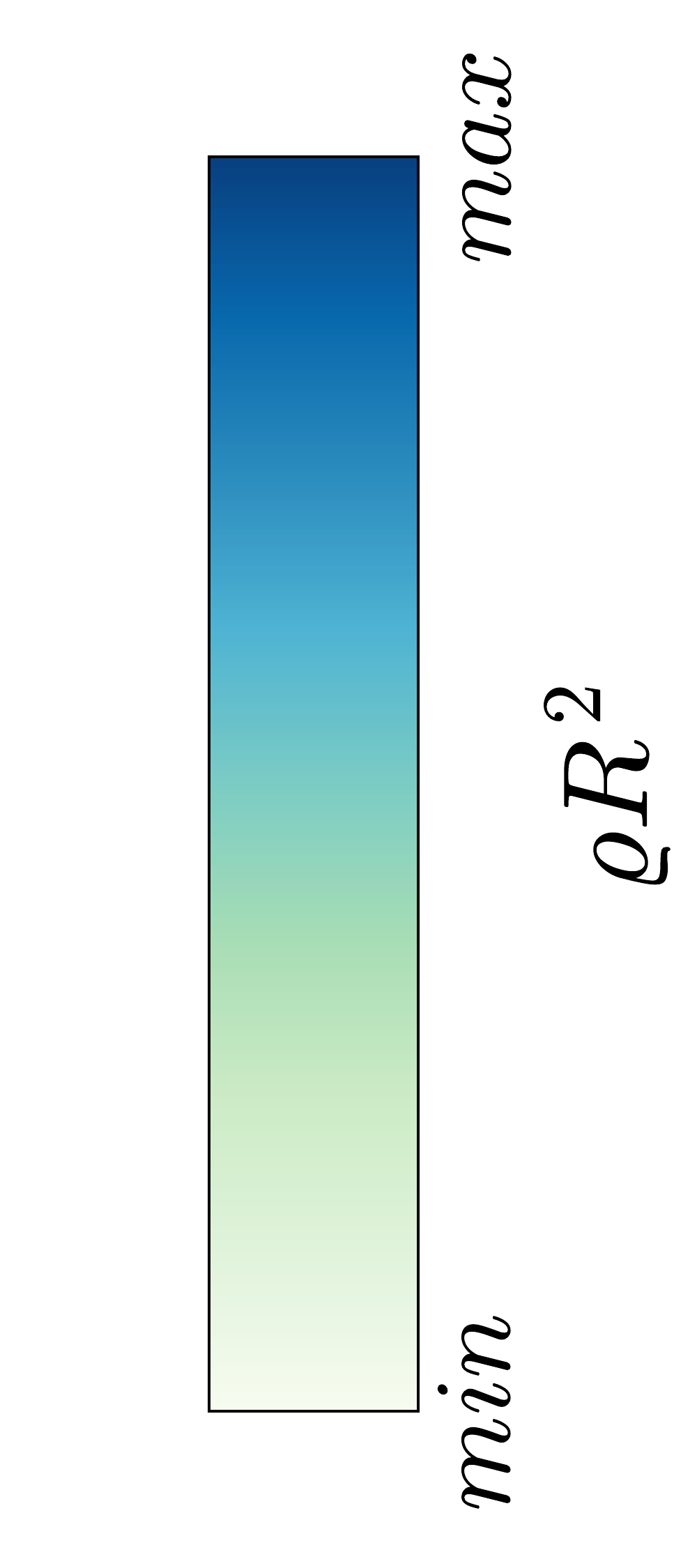}
\caption{A selection of cluster crystals configurations. For each configuration we show (from left to right) a snapshot of the MC simulation, the density profile obtained from DFT, and the Voronoi representation of the distribution. To help visualize the DFT profiles, alongside the color map, we display clusters as spikes whose height is proportional to the local density. In the Voronoi representation, we classify each cluster according to its topological charge, as shown in the legend.}
\label{clus_snap}
\end{figure*}

\subsection{Description of the cluster crystals}
\label{crystals}

In the previous section we have determined the phase diagram of the model, observing that, as for bulk systems, cluster crystals can be observed at high densities. It is then interesting to characterize the features of these crystals.

In figure \ref{clus_snap} we show a selection of the observed cluster crystal configurations, comparing snapshots of the MC simulations and the density profile predicted by DFT. We also compute the Voronoi representations of the density profiles by partitioning the sphere surface according to the distance from the center of mass of each cluster. Each cell contains all the points which are closer to the center of the cluster corresponding to that cell than to any other center. This partitioning reveals one of the staple features of the spherical cluster crystals, that is the presence of defective lattice sites with a coordination number lower than that expected for hexagonal lattices observed in bulk systems \cite{Prestipino_2014}.

This is not the result of a clumsy minimization. In fact, these defects are the result of topological frustration, i.e., the impossibility to propagate a favored local order over the whole space. The geometrical constraints given by the sphere surface lead to the formation of ineliminable topological defects called disclinations, corresponding to lattice sites which have an higher or, in our case, lower coordination number than that expected for systems which are not frustrated.

The total number of disclinations is strictly connected to the topology of the sphere. One can see this by using the Euler characteristic, which states that for a convex polyhedron, such as a Voronoi diagram on a sphere surface, one has

\begin{figure}[b!]
\captionsetup[subfigure]{labelformat=empty}
\includegraphics[width = 3.5cm]{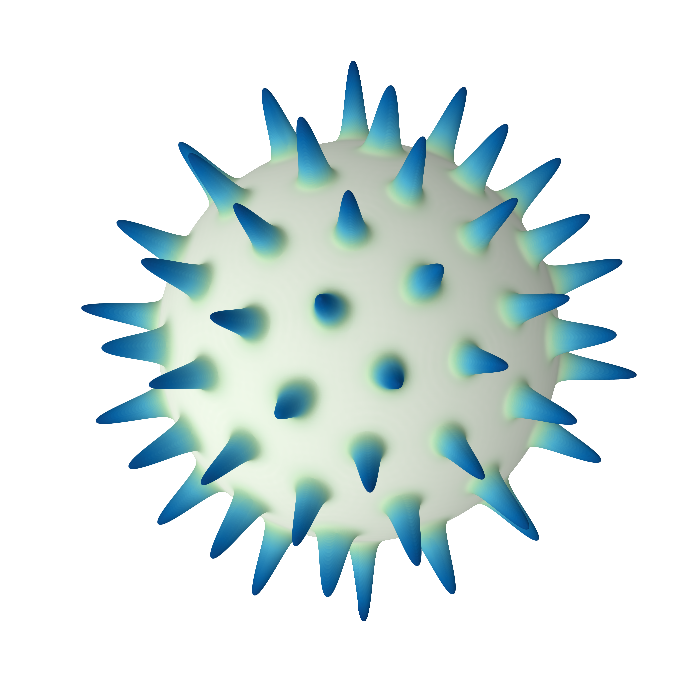}
\includegraphics[width = 3.5cm]{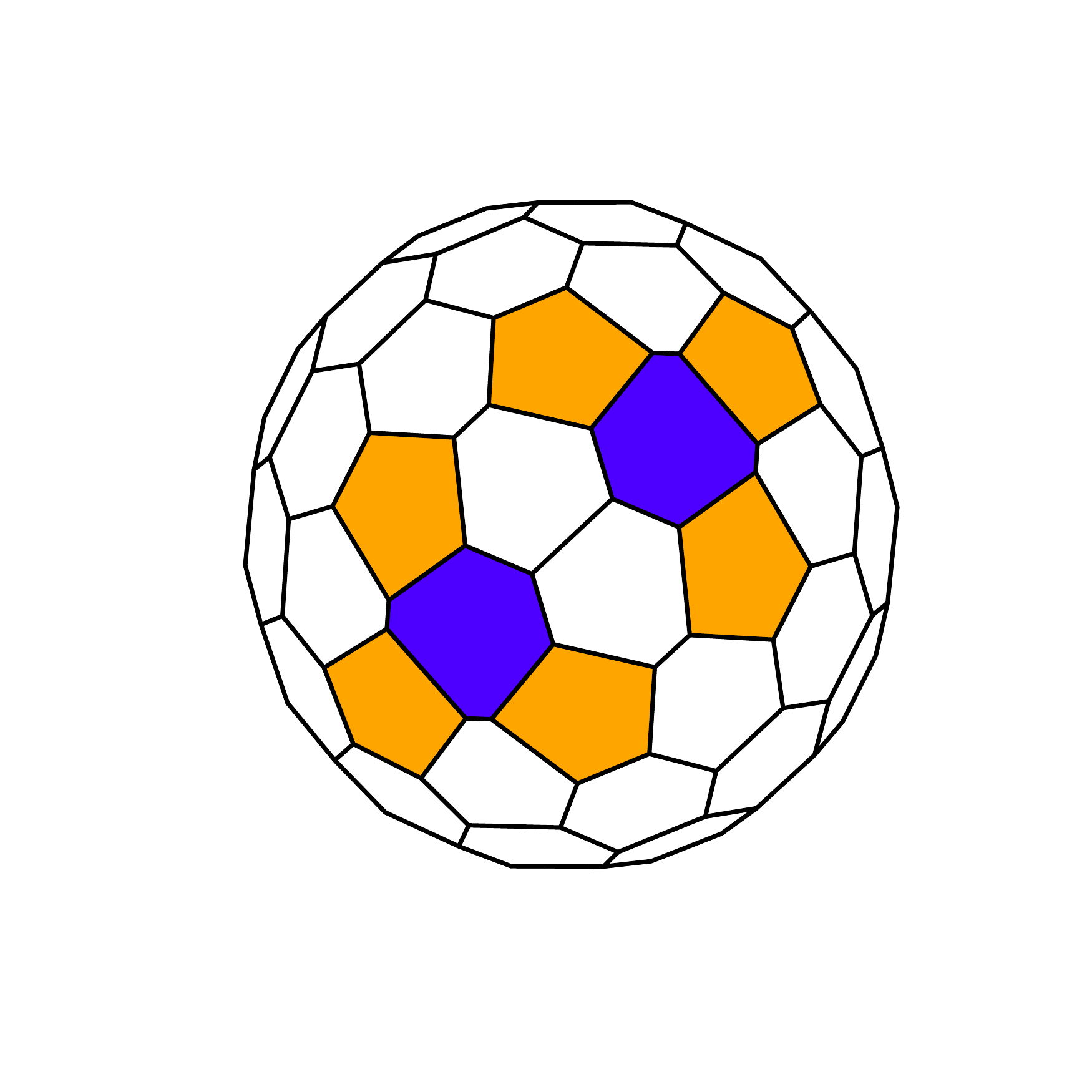}
 \includegraphics[width = 1.5cm]{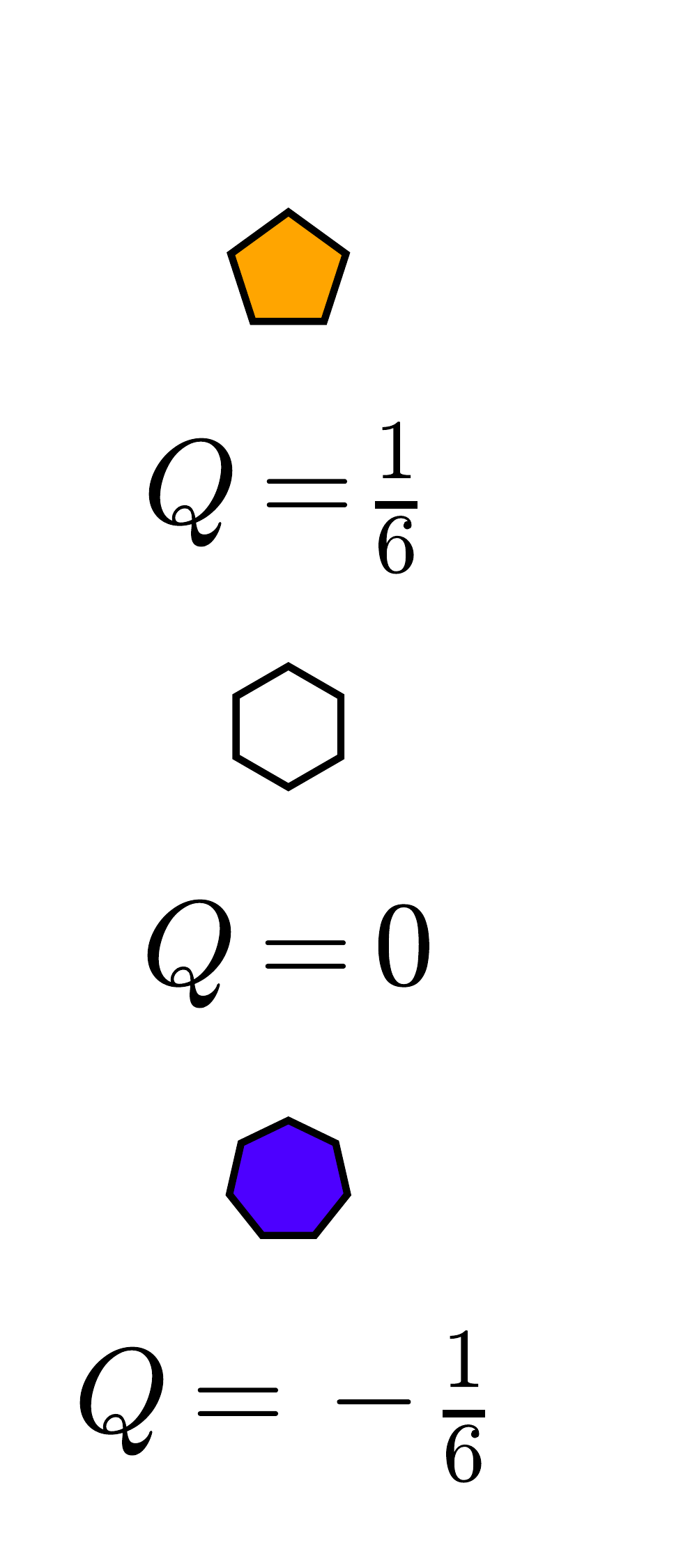}
\caption{A metastable configuration in which sevenfold disclinations can be observed. Positively charged disclinations tend to aggregate near the sevenfold disclinations, forming dislocation defects, in order to screen their negative topological charge.}
\label{dislocation}
\end{figure}

\begin{equation}
	V - E + F = 2
\end{equation}

where $V$ is the number of vertices, $E$ is the number of edges, and $F$ is the number of faces of the polyhedron. Then if we suppose that all faces are either hexagons or pentagons, it is easy to see that we need exactly $12$ pentagons to satisfy this geometrical constraint, while the number of hexagons is free to vary.

Another way of obtaining the same constraint is to assign to each defect a topological charge, which is proportional to the difference between the coordination number in the regular lattice and the coordination number of the disclination. So for example clusters with five neighbors have a topological charge $Q = 1/6$, while clusters with seven neighbors have $Q = -1/6$ \cite{Nieves_2016}. Moreover the total topological charge on a surface is associated with its Gaussian curvature $G$ ($1/R^2$ for the sphere) by the relation \cite{Nieves_2016}

\begin{equation}\label{topological}
	Q_{tot} = \frac{1}{2\pi}\int dS\,G = 2.
\end{equation}

Disclinations with seven or more neighbors have negative topological charge. They can still be encountered in some metastable configurations, as seen in figure \ref{dislocation}, but each of these defects must be countered by an additional disclination of opposite charge, leading to an even more frustrated configuration, which may explain why these configurations are never the most stable ones. Indeed, as shown again in figure \ref{dislocation}, negatively charged disclinations are coupled to positively charged ones, forming dislocations. This is a result of the fact that the Gaussian curvature of the sphere is positive and constant, so the condition in equation \eqref{topological} must be enforced locally, requiring positively and negatively charged disclinations to screen each other, otherwise the sphere would display a local indentation.

Since the cluster crystals contain topologically distinct clusters, in order to study the characteristics of these structures we distinguish between clusters with different topological charges and compare them.

\begin{figure}[t]
\includegraphics[width = 3in]{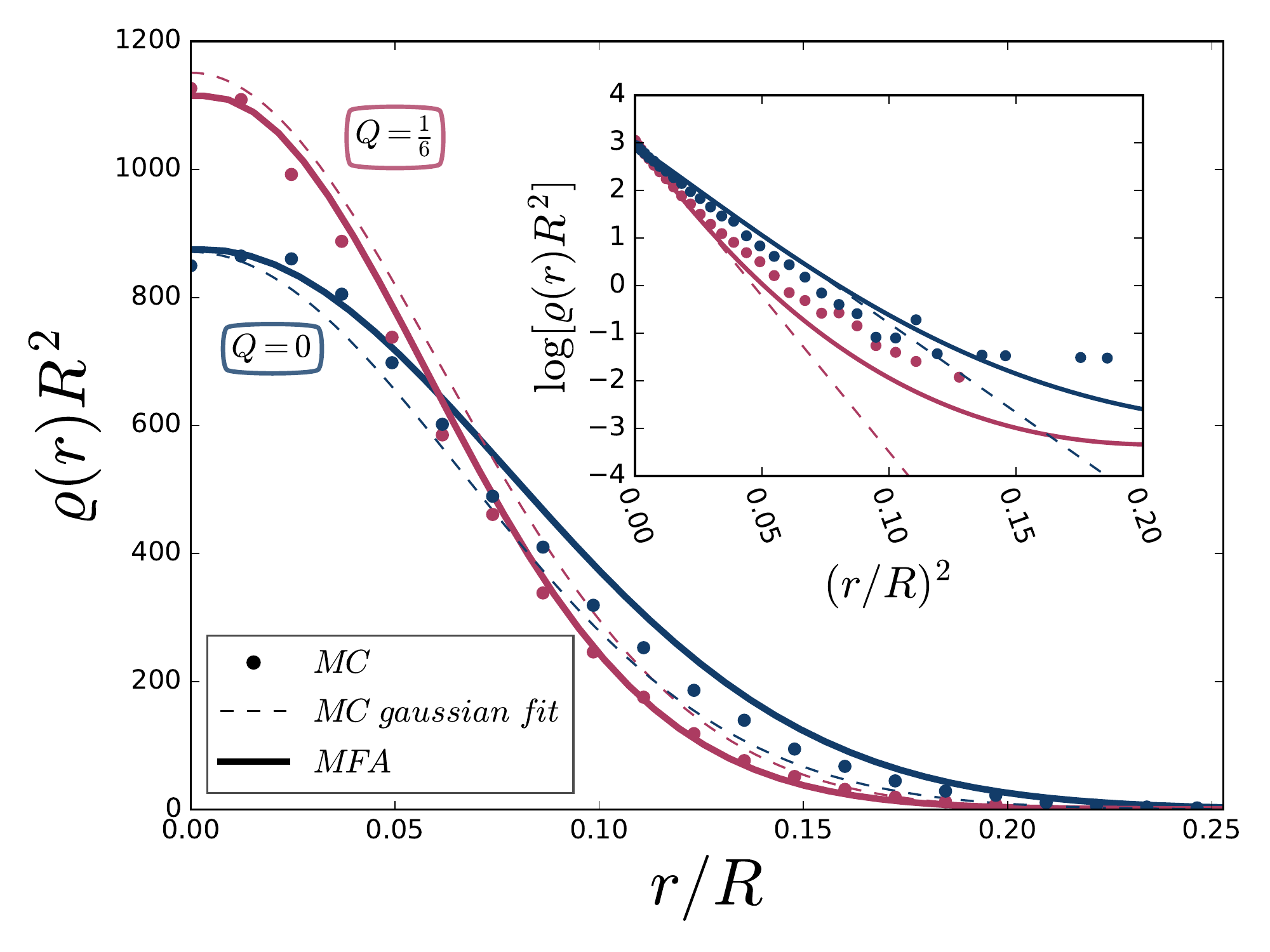}
\caption{Profile of a cluster at $T^*=1$, $\sigma/R=0.68$ and $\varrho R^2 \sim 35.0$ sampled along the line connecting it to one of its neighbors. Results from the MC simulation are compared to DFT results and a Gaussian fit. The semi-logarithmic inset shows that the Gaussian approximation is only valid near the cluster center. }
\label{clus_prof}
\end{figure}

We begin by looking at the density profiles of the clusters, plotted in figure \ref{clus_prof}. For both defective and non-defective clusters, DFT as well as MC results resemble a Gaussian, as shown in the semi-logartithmic inset, where deviations from the Gaussian shape at large distances from the center of the cluster are also evidenced. Defective clusters tend to be smaller but with an higher maximum density than the non-defective clusters. 

To understand the energetic effect of the disclinations on the system, we first define a local grand potential $\beta\omega_D$ which is the grand potential of a single cell of our grid

\begin{equation}
\begin{aligned}
	\beta\omega_D(\bx) = \frac{\pi^2}{N^2}\sin\theta_{\bx}\subs{\varrho}{x}
	\big[ \ln \big( \subs{\varrho}{x}/\brho\big) -1 - \beta\mu^{ex}\big]\\
	+ \frac{\pi^4\beta}{2N^4} \sin \subs{\theta}{x} \subs{\varrho}{x} \sum_{\bx'} \sin \theta_{\bx'} 
	\varrho_{\bx'} w_{\bx,\bx'}
\end{aligned}
\end{equation}

\begin{figure}[t]
\includegraphics[width = 3in]{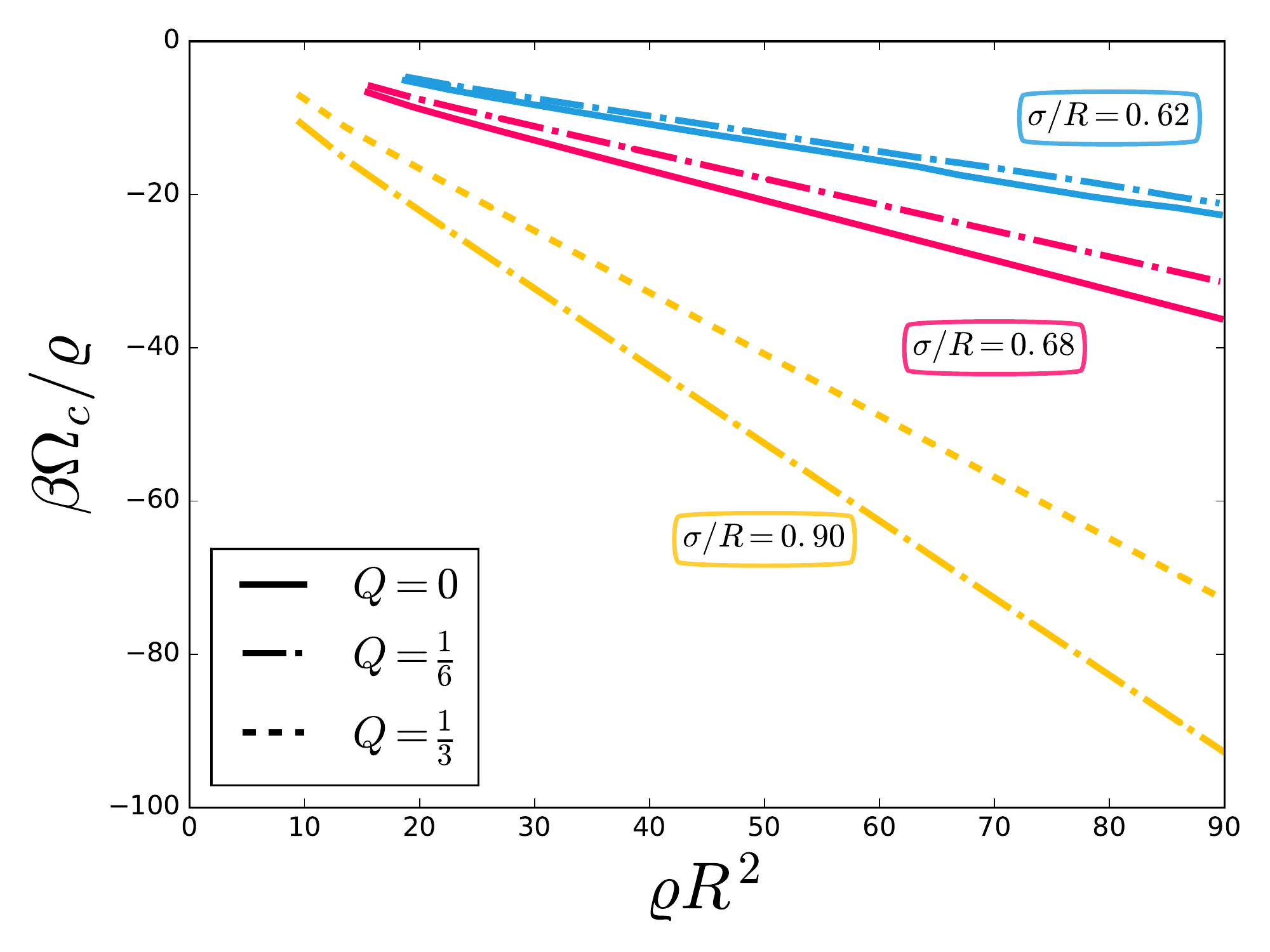}
\caption{The local grand potential per cluster plotted as a function of the mean density of the system for different configurations, at $T^* = 1$. We distinguish between clusters with different topological charges, finding that defective clusters are less stable than their non defective counterparts.}
\label{clus_gpot}
\end{figure}

The sum of $\beta\omega_D(\bx)$ over the cells of a cluster define its local grand potential $\beta\Omega_{c}$, which can be used to compare its stability to that of other clusters. In figure \ref{clus_gpot} we compare the local grand potentials $\beta\Omega_c$ of clusters with different topological charges. 

Clusters sitting on a disclination are less stable than non defective clusters of the same distribution. Moreover, the energy penalty increases as the topological charge of the disclination increases, so the formation of disclinations with larger topological charges is disfavored with respect to fivefold disclinations. This explains why stable configurations do not contain fourfold or higher-order disclinations, unless they are forced to by the geometrical constraints when the ratio $\sigma/R$ becomes too large to allow, at least, a stable $12$-cluster configuration, as seen in figure \ref{clus_snap} for the $10$-cluster configuration.

\begin{figure}[h!]
\includegraphics[width = 3in]{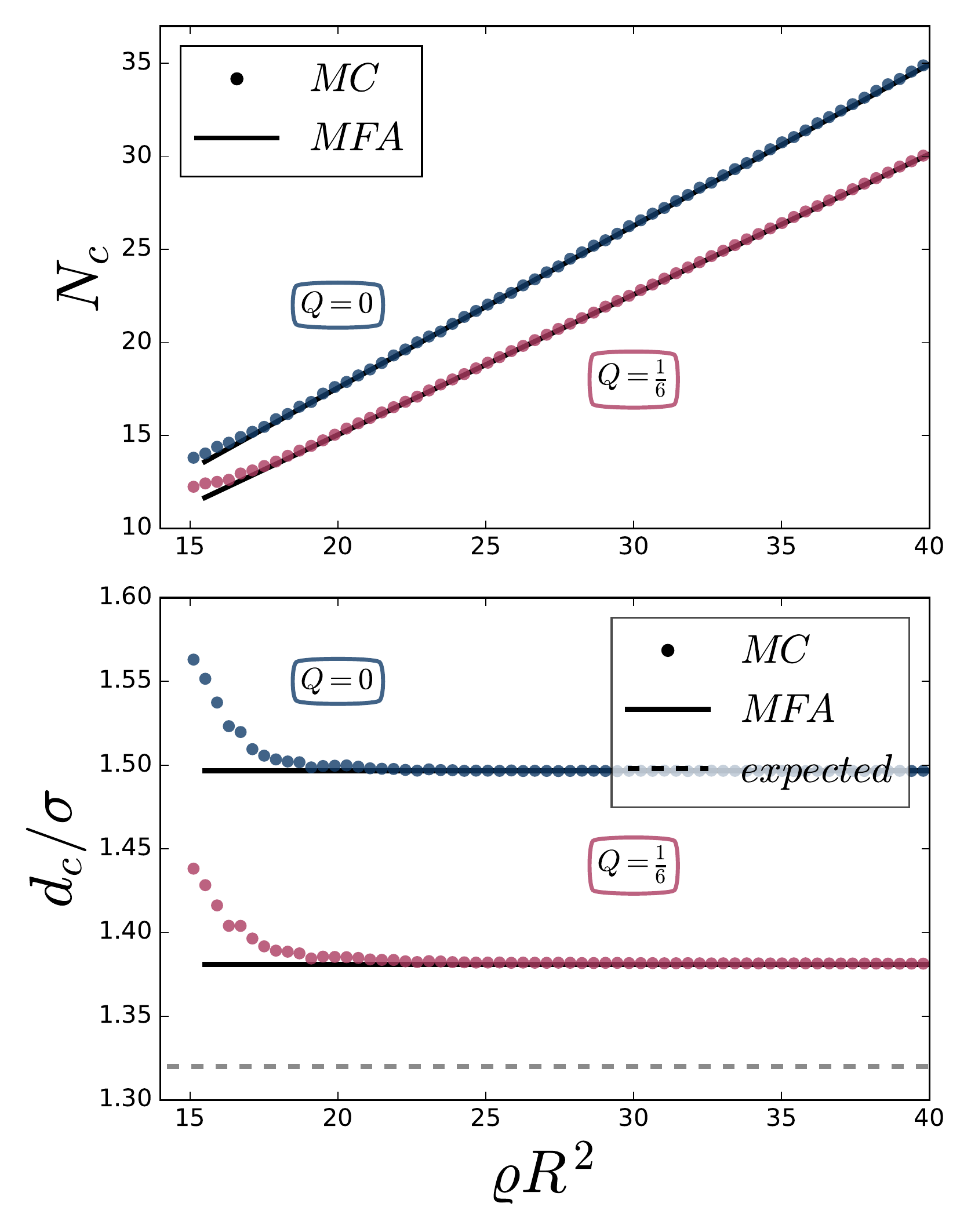}
\caption{Some of the cluster characteristics which are influenced by the topological charge. We plot, at $T^*=1$, the number of particles per cluster $N_c$ as a function of the mean density in the upper panel, and the mean distance between neighboring clusters $d_c/\sigma$ in the lower. For clarity we only show the data from the 16-cluster configuration at $\sigma/R = 0.68$. For each plot, we compare the DFT predictions with the MC simulations results, showing excellent agreement between the two at large density values. The dotted line in the lower panel refers to the estimate $d_{c}=2\pi R/\ell^{*}$. }
\label{clus_char}
\end{figure}

The stability is not the only characteristic of the clusters which is affected by their topological charges. In figure \ref{clus_char} we plot the number of particles contained in each cluster $N_{c}$ (upper graph) and the distance from neighbors $d_{c}$ (lower graph) as a function of the mean density. The DFT predictions for these quantities display excellent agreement with the results of the MC simulations, aside from an initial discrepancy due to the crossover of the simulated system from the fluid to the cluster crystal. We remark, however, that the obtained distances are larger than the expected $d_{c} = \tfrac{2\pi R}{\ell^*}$.

We also observe that, as for the bulk systems \cite{Pini_2015, Mladek_2007}, the number of particles per cluster grows linearly with the mean density $\varrho$, while the distance between neighbors is nearly independent from the density. These are different aspects connected to the resilience of cluster crystals to compression \cite{Likos_2001}: in the bulk, the cluster crystal lattice constant does not change upon compression. It is instead more convenient for the system to redistribute the particles into existing clusters. Notice that in the spherical case it is not equivalent to change the volume or the number of particles of the fluid \cite{Tarjus_2011}, because of the finite size of the system. Nevertheless, the cluster crystals exhibit the same behavior as in the bulk when the mean density is increased by increasing the number of particles. 

The linear scaling of $N_{c}$ with $\varrho$ is also related to the linear behavior of $\beta\Omega_{c}/\varrho$ as a function of $\varrho$ displayed in figure \ref{clus_gpot} for both vanishing and non-vanishing $Q$. Each particle on a cluster interacts with $N_{c}-1$ particles on the same cluster and with $N_{c}$ particles on different clusters, so that the energy per particle depends linearly on $N_{c}$. If $\Omega$ is dominated by its energetic contribution, one then expects $\beta\Omega_{c}/N_{c}$ to be a linear function of $N_{c}$. Since $N_{c}\sim\varrho$, the same holds for $\beta\Omega_{c}/\rho$ as a function of $\rho$.

Here, however, we can also compare the behavior of clusters with different topological charges. We notice that defective clusters tend to be closer to their neighbors and contain fewer particles.

\begin{figure}[t]
\centering
\includegraphics[width = 2.8in]{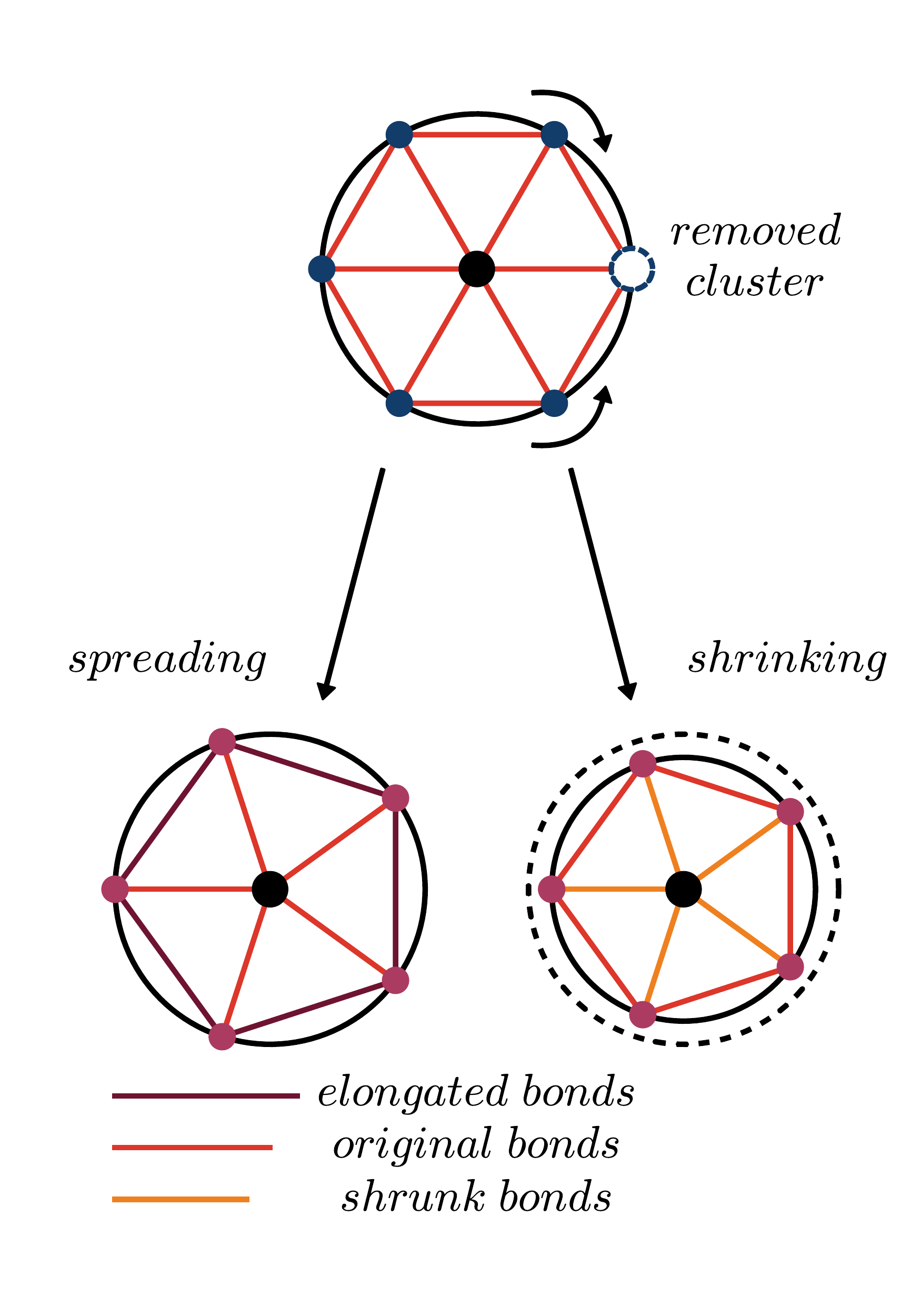}
\caption{Possible relaxation mechanisms of the first neighbors shell of a disclination. If a cluster is removed from the shell, the others take a pentagonal configuration, either keeping a fixed distance from the center by spreading around it, or keeping the distance between each other fixed by shrinking around the center.}
\label{clus_exp}
\end{figure}

The former property can be explained by geometrical considerations. If one removes one of the neighbors of a non-defective cluster, the remaining neighbors will take a pentagonal arrangement. They have two ways of doing so: either they maintain the original distance from the central cluster, but grow farther from each other in the process, which we call spreading mechanism, or they maintain their original distance from each other, but come closer to the central cluster, which we call shrinking mechanism.
We illustrate the two mechanisms in figure \ref{clus_exp}.

We observe that the spreading mechanism does not take place in cluster crystals. Indeed, one may argue that this mechanism is unfavorable to the overall stability of the crystal, because by creating larger gaps between the remaining clusters of the neighbor shell, it allows the particles of each cluster to spread under the push of the internal repulsion, disrupting the cluster. Hence, we argue that frustration is released only on the central cluster by way of the shrinking mechanism.

\begin{figure}[t]
\includegraphics[width = 3in]{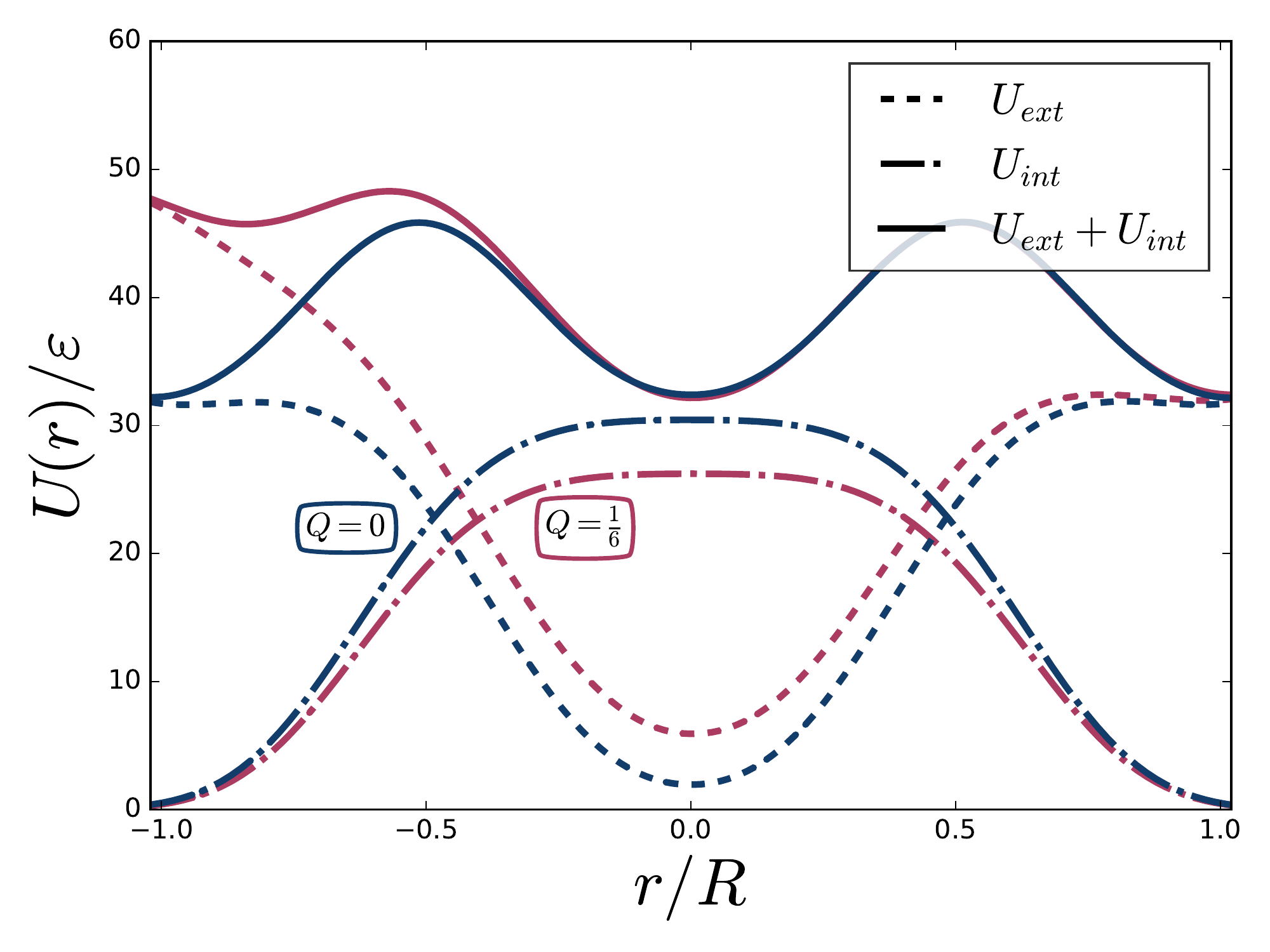}
\caption{Different contributions to the potential energy landscape in proximity of clusters with six neighbors (blue lines) and five neighbors (purple lines): we distinguish the potential $U_{ext}$ produced by the surrounding clusters (dashed lines) from the internal repulsion $U_{int}$ between particles of the same cluster (dash-dotted lines). The sum of the two produces identical minima of the potential energy landscape irrespective of the topological charge of the clusters. The potentials are sampled along the great circle connecting the central clusters to one of their neighbors.}
\label{clus_stab}
\end{figure}

This also affects the number of particles contained in the disclination site. To understand why disclinations contain fewer particles than non-defective clusters, we compare in figure \ref{clus_stab} the energy landscapes generated by the cluster configuration in the vicinity of clusters with different topological charges. We find that, because of their closeness, the inter-cluster energy minimum generated by the neighbors of the disclination is shallower than the one found for non-defective clusters, and as a result the number of particles the site can host is smaller.

In fact the minima in the total energy landscape, given by the sum of the inter cluster and intra cluster terms, must be equivalent for all clusters at equilibrium, otherwise we would observe a net flow of particles from the shallower minima to the deeper ones. Thus, by containing fewer particles which contribute to intra cluster repulsions, disclinations compensate for the unfavorable inter cluster interactions.

\section{Conclusions}
\label{conclusions}

In this work we have explored the clustering of ultra-soft particles on the surface of a spherical substrate. 

Using the same ideas developed in ref. \citet{Likos_2001}, we have extended the clustering criterion to fluids on spherical surfaces. We found that a negative minimum in the spherical harmonic expansion of the pair potential at an harmonic degree $\ell^* \neq 0$ leads to clustering at high enough densities. 

We focused our attention on a specific model, the GEM-4 fluid, using DFT to obtain its phase diagram. In order to solve the DFT equations we employed a version of the minimization algorithm developed in ref. \cite{Pini_2015} generalized to the spherical system, which allows us not to make \textit{a priori} assumptions about the functional form of the density profile. We also performed Monte Carlo simulations to test the theory and assess the full effect of correlations in the fluid, and found quantitative agreement between theory and simulations both in the low-density fluid phase, and in the high-density cluster crystal phases. 

Similarly to the bulk three-dimensional case \cite{Mladek_2006, Kahl_2007}, cluster crystals are predicted to form at all temperatures, provided the density is sufficiently high. However, the finite size of the system introduces an additional degree of freedom, namely, the ratio $\sigma/R$ between the size of the particles and the radius of the sphere. Indeed, the number of clusters of the crystal configurations is determined by this ratio, while being nearly independent from either density or temperature. Each configuration spans a certain interval of $\sigma/R$ and, as $\sigma/R$ is decreased, the number of clusters of the more stable configuration increases. 

We also studied the effect of topological frustration induced on these structures by the sphere curvature. Ineliminable disclination defects in the crystal lattice display many differences from non-defective sites: they are less stable, contain fewer particles, and their neighbors are closer to them. We explain the latter property as a geometrical effect, and find that because of the reduced distance between the disclination and its neighbors, the inter-cluster repulsions form a shallower minimum in the energy landscape at the disclination site, so that the cluster can only contain a reduced number of particles.

We remark that this work is far from exhaustive and many interesting aspects of the system could be the topic of future developments.

For example, we have not explored how the curvature of the sphere influcences the dynamical properties of the GEM-4 fluid, such as the nucleation process\cite{Nikoubashman_2012} or the diffusivity of single particles in the cluster crystal phase \cite{Montes_Saralegui_2013, Coslovich_2011, Moreno_2007}.

Another interesting topic is the role of the substrate in clustering. Recent works show that a coupling between the density field and the local curvature of the sphere, by way of capillary interactions, can lead to microphase separation and patterning \cite{Lavrentovich_2016, Marenduzzo_2013, Cacciuto_2013}. A more realistic model for lipid membranes covered by soft particles would then need to take the fluctuations of the substrate shape into account.

Moreover, the attitude of these soft particles to form patches on the surface of a spherical substrate naturally suggests their exploitation in the design and manufacture of patchy particles \cite{Yi_2013, Bianchi_2011}. In order to fully take advantage of the properties of these systems, a multiscale modeling approach is needed to understand how to tune the hierarchical self-assembly of soft particles into patches, and patchy particles into mesoscopic structures of interest. 

\section*{Conflicts of interest}

There are no conflicts to declare.

\section*{Acknowledgements}
One of the authors (S. F.) wishes to thank the \textit{Laboratorio di Calcolo e Multimedia} (LCM) for providing the computers used for the numerical parts of this work.

\section*{Appendix A: DETAILS OF THE MINIMIZATION ALGORITHM}
\label{appendix}

The algorithm employed here is adapted from the algorithm presented in reference \cite{Pini_2015}, which is itself based on the steepest descent algorithm, where the equilibrium density profile $\varrho(\x)$ is obtained recursively as

\begin{equation}\label{steep}
	\varrho(\x)^{(k+1)} = \varrho(\x)^{(k)} - \delta \frac{\delta \beta\Omega[\varrho]}{\delta\varrho(\x)}\Bigm|_k.
\end{equation}

Where $k$ refers to the $k$-th iteration of the process. Naturally, we need to discretize the density profile $\varrho(\x)$, as well as the grand potential functional $\beta \Omega[\varrho]$, over a grid of $K \times 2K$ points in order to treat the problem numerically, as illustrated in section \ref{DFT}. 

Doing so requires some care in the passage from the functional differentiation with respect to the density profile $\varrho(\x)$ to the differentiation with respect to the discrete sampled densities $\varrho_{x}$: to go from the former to the latter we need to divide by the area of the sampling lattice cell

\begin{equation}
	\frac{\delta}{\delta\varrho(\x)} \longmapsto \frac{K^2}{\pi^2\sin \theta_{\bx}}\frac{\partial}{\partial\varrho_{\bx}}
	\equiv \bpar_{\varrho_{\bx}}.
\end{equation}

In fact by defining the partial derivative this way, we can obtain the discretized version of equation \eqref{steep} either by taking the functional derivative of the continuous grand potential and then discretizing the result, or by discretizing the grand potential and then taking its partial derivative.

The needed partial derivative $\partial\beta\Omega_D\{\varrho_{\bx}\}/\partial\varrho_{\bx}$ of the discretized functional $\beta\Omega_D$ is given by the left-hand side of equation \eqref{EL_D}

\begin{equation}
	\frac{\partial\beta\Omega_D\{\varrho_{\bx}\}}{\partial\varrho_{\bx}} = \ln\big( \subs{\varrho}{x}/\brho\big) - \beta\mu^{ex} + 
	\beta \sum_{\ell,m} \sqrt{\tfrac{4\pi}{2\ell+1}}w_{\ell}\varrho_{\ell,m}Y_{\ell,m}(\bx)
\end{equation}

Here we have used the convolution theorem to deal with the double summation. In the case of spherical harmonic expansions, the convolution theorem for continuous functions takes the form \cite{Driscoll_1994}

\begin{equation}
\begin{aligned}
	\int_{S^2} \dx'\,\varrho(\x')w(\x',\x)	= \sum_{\ell,m} \sqrt{\tfrac{4\pi}{2\ell+1}}w_{\ell}\varrho_{\ell,m}Y_{\ell,m}(\x)\\
	\int\int_{S^2}\dx\dx'\,\varrho(\x)\varrho(\x')w(\x',\x) = \sum_{\ell,m} \sqrt{\tfrac{4\pi}{2\ell+1}}w_{\ell}|\varrho_{\ell,m}|^2.
\end{aligned}
\end{equation}

However, in our case we consider discretized densities $\varrho_{\bx}$ and interaction potentials $w_{\bx,\bx'}$. As for the Fourier transform, the discretization in direct space introduces a cutoff in the spherical harmonic expansion $\ell_{max} = \tfrac{1}{2}N -1$ \cite{Driscoll_1994}.

Fast transform algorithms are available to obtain the harmonic coefficients from a sampled function and viceversa. Here we employ the SHTOOLS package \cite{SHTOOLS}, which is freely available online, based on the FFTW3 package \cite{Frigo_2005}.

As for the algorithm in \cite{Pini_2015}, we make several improvements with respect to the steepest descent algorithm to achieve faster convergence.

The first step is to use the Jacobi preconditioner, which means that we replace equation \eqref{steep} with

\begin{equation}\label{precon}
	\varrho_{\bx}^{(k+1)} = \varrho_{\bx}^{(k)} - \delta \xi_{\bx}^{(k)},
\end{equation}

where the quantities $\xi_{\bx}^{(k)}$ are defined as

\begin{equation}\label{def_precon}
	\xi_{\bx}^{(k)} = \big( \bpar_{\varrho_{\bx}} \beta\Omega_D |_{k} \big)\times\big( \bpar^2_{\varrho_{\bx}} \beta\Omega_D |_{k}\big)^{-1}
\end{equation}

where the second derivative $\bpar_{\varrho_{\bx}}^2 \beta\Omega_D$ is given by

\begin{equation}\label{ddomega}
	\big(\bpar_{\varrho_{\bx}}^2 \beta\Omega_D \big)^{-1}= \frac{\pi^2 \sin \theta_{\bx}}{K^2}\frac{\varrho_{\bx}}{1 + \tfrac{\beta\pi^2}{K^2}\sin\theta_{\bx}\varrho_{\bx}w(0)}
\end{equation}

The purpose of this modification is to alter the shape of the elongated basins of attraction in the grand potential landscape, in order to obtain more circular basins, which are better suited to be treated with this algorithm.

However here we see that the measure factor $\tfrac{\pi^2}{K^2}\sin\theta_{\bx}$ in the derivative leads to a null preconditioner $\xi_{\bx}$ at the poles, which is clearly a mistake, since it would mean that the density at the poles would not change from its initial value. The reason for this problem is the fact that, since the spacing in the grid is finite, the correct measure of each cell should be proportional to $\cos(\theta_{\bx}) - \cos(\theta_{\bx}+\Delta\theta)$, instead of $\sin\theta_{\bx}$. In principle we could make this substitution in equation \eqref{ddomega} to obtain a formally correct preconditioner, however we find that it is more stable to simply drop the measure factor and redefine the preconditioner as

\begin{equation}
		\xi_{\bx}^{(k)} = \frac{K^2}{\pi^2\sin\theta_{\bx}}\big( \bpar_{\varrho_{\bx}} \beta\Omega_D |_{k} \big)\times\big( \bpar^2_{\varrho_{\bx}} \beta\Omega_D |_{k}\big)^{-1}
\end{equation}

Moreover we follow the conjugate gradient method and actually determine the descent direction by a linear combination of the preconditioners at each step, so that the recursive relation becomes

\begin{equation}\label{conj}
	\varrho_{\bx}^{(k+1)} = \varrho_{\bx}^{(k)} - \delta \psi_{\bx}^{(k)},
\end{equation}

where $\psi_{\bx}$ is obtained through the following relations

\begin{equation}
	\begin{aligned}
	\psi_{\bx}^{(k)} = \xi_{\bx}^{(k)} + \zeta^{(k)}\psi_{\bx}^{(k-1)}\\
	\zeta^{(k)} = \frac{\sum_\bx \xi_{\bx}^{(k)}\big(\xi_{\bx}^{(k)} - \xi_{\bx}^{(k-1)}\big)  }{\sum_{\bx} \big( \xi_{\bx}^{(k-1)} \big)^{2}}
	\end{aligned}
\end{equation}

As in ref. \cite{Pini_2015} we use a preconditioned steepest descent step every 10 step of the preconditioned conjugated gradients algorithm, in order to improve the stability of our minimization algorithm.

The sampled densities $\varrho_{\bx}$ were updated according to equations \eqref{conj} until the partial derivatives $\bpar_{\varrho_{\bx}} \beta\Omega_{D}$ vanish within a prescribed accuracy. More specifically the iteration was stopped when $\sum_{\bx} \big( \bpar_{\varrho_{\bx}} \beta\Omega_{D} \big)^{2}$ became smaller than $10^{-15}$.

The step size $\delta$ must be chosen with care to achieve convergence in a sensible time. To obtain the optimal choice for $\delta$ one should minimize the function $g^{(k)}(\delta) = \beta\Omega_{D}(\{\varrho_{\bx}^{(k)} - \delta \psi_{\bx}^{(k)}\})$, which must be thought of as a function of $\delta$ at fixed $\varrho_{\bx}^{(k)}$ and $\psi_{\bx}^{(k)}$.

As this does not yield an explicit expression for $\delta$, its value is obtained here through a single step of the Raphson-Newton method, which amounts to setting

\begin{equation}
	\delta^{(k)} = - \partial_{\delta} g^{(k)} \big( \partial_{\delta}^2 g^{(k)}\big)^{-1}
\end{equation}

where the derivatives are given by

\begin{equation}\label{dg}
	\partial_{\delta}g^{(k)} = - \frac{\pi^2}{K^2}\sum_{\bx} \sin\theta_{bx} \bpar_{\varrho_{\bx}}\beta\Omega_{D} \psi_{\bx}^{(k)}
\end{equation}
\begin{equation}\label{ddg}
	\partial^2_{\delta}g^{(k)} = \frac{\pi^4}{K^4}\sum_{\bx,\bx'} \sin\theta_{\bx}\sin\theta_{\bx'} \frac{\bpar^2\beta\Omega_D}{\partial\varrho_{\bx}\partial\varrho_{\bx'}}\psi_{\bx}^{(k)}\psi_{\bx'}^{(k)}
\end{equation}

The quantities in equation \eqref{dg} are readily obtained by combining the results above. To evaluate equation \eqref{ddg} we need to use the convolution theorem

\begin{equation}
	\partial^2_{\delta}g^{(k)} = \frac{\pi^2}{K^2}\sum_{\bx} \sin\theta_{bx} \frac{\big(\psi_{\bx}^{(k)}\big)^{2}}{\varrho_{\bx}} + \beta\sum_{\ell,m} \sqrt{\frac{4\pi}{2\ell+1}}\big(\psi_{\ell,m}^{(k)}\big)^2 w_{\ell}
\end{equation}

Since the grand potential functional is not globally convex, the optimization described by equations \eqref{dg}, \eqref{ddg} can also lead to a negative step-size. In order to avoid such an occurrence we set a fixed positive value of $\delta$, which we use whenever the optimization would give a negative value.

Moreover we note that the densities must obviously be non-negative, so whenever a minimization step would lead to a negative value of the local density, we replace that density with a small positive value ($\sim 10^{-15}$ in this work). Of course null densities are perfectly acceptable, and in principle one could think that the equilibrium density profile could also contain some null local densities. However for that to be the case, the value of $\bpar_{\varrho_{\bx}}\beta\Omega_D$ evaluated at $\varrho_{\bx}=0$ must be non-negative, which is easily shown not to be possible because of the logarithmic divergence of the derivative at null density.



\balance

\renewcommand\refname{References}


\nocite{HUMP96}
\nocite{STON1998}

\bibliography{biblio_sft} 

\bibliographystyle{rsc} 

\end{document}